\documentclass[aps,pra,twocolumn,showpacs,superscriptaddress,floatfix,10pt]{revtex4-2}
\usepackage{framed}
\usepackage{graphicx}
\usepackage{xfrac}
\usepackage{amsmath}
\usepackage{epsfig}
\usepackage{helvet}
\usepackage{amssymb}
\usepackage{soul}
\usepackage{xcolor}
\usepackage{framed}
\usepackage{hyperref}
\usepackage{amsthm}
\usepackage{esint}
\usepackage{mathtools}

\usepackage[scr=boondoxupr]{mathalpha}
\newcommand*{\rightangle}{%
    \mathchoice%
        {\mathrel{\makebox[7pt][c]{\rule{.4pt}{7.5pt}\rule{5pt}{.4pt}}}}%
        {\mathrel{\makebox[7pt][c]{\rule{.4pt}{7.5pt}\rule{5pt}{.4pt}}}}%
        {\mathrel{\makebox[5.5pt][c]{\rule{.4pt}{5.25pt}\rule{3.5pt}{.4pt}}}}%
        {\mathrel{\makebox[4pt][c]{\rule{.4pt}{3.75pt}\rule{2.5pt}{.4pt}}}}%
}

\def\myfloor#1{\left\lfloor {#1} \right \rfloor}

\begin{document}
\title{Holographic Entropy Inequalities and the Topology of Entanglement Wedge Nesting}
\author{Bart{\l}omiej Czech, Sirui Shuai, Yixu Wang and Daiming Zhang}
\affiliation{Institute for Advanced Study, Tsinghua University, Beijing 100084, China}
\vskip 0.15cm

\begin{abstract}
\noindent
We prove two new infinite families of holographic entropy inequalities. A key tool is a graphical arrangement of terms of inequalities, which is based on entanglement wedge nesting (EWN). It associates the inequalities with tessellations of the torus and the projective plane, which reflect a certain topological aspect of EWN. The inequalities prove a prior conjecture about the structure of the holographic entropy cone and show an interesting interplay with differential entropy.
\end{abstract}

\maketitle

\textit{Introduction.---} Recent years have revealed the importance of classifying and studying quantum states according to their patterns of entanglement. One important class of quantum states are those whose entanglement entropies can be computed by a minimal cut prescription. The prescription assumes that the state can be represented by an auxiliary `bulk' structure, typically a tensor network or---in holographic duality \cite{adscft}---a bulk geometry. The minimal cut prescription equates the entanglement entropy of a region $X$ with the weight of a bulk cut, which separates $X$ from $\overline{X}$ (the complement of $X$). The prescription is valid for all random tensor network (RTN) states \cite{rtn} at large bond dimension and---in holographic duality---for the dominant area term in the Ryu-Takayanagi proposal \cite{rt1, rt2, qes}.

This paper concerns constraints on entanglement entropies, which are implied by the minimal cut prescription. Because of the application to holographic duality, such constraints are conventionally called `holographic entropy inequalities.' In the vector space of hypothetical assignments of entropies to regions (entropy space), the locus of saturation of each holographic inequality is a hyperplane. Consequently, the set of entropies allowed by all holographic inequalities is called the `holographic entropy cone' \cite{hec}. Further following the holographic nomenclature, we will refer to weights of cuts as `areas.' 

The simplest holographic inequality, known as the monogamy of mutual information \cite{mmiref}, is:
\begin{equation}
S_{AB} + S_{BC} + S_{CA} \geq S_A + S_B + S_C + S_{ABC}
\label{mmi}
\end{equation}
Here $A,B,C$ are disjoint subsystems and $S_{X}$ are their entanglement entropies. Union signs for disjoint regions are implied, for example $AB \equiv A \cup B$. Other than (\ref{mmi}), currently known holographic entropy inequalities include one single-parameter infinite family (\ref{cyclic}) \cite{hec} and 378 other isolated inequalities \cite{cuenca, withyunfei, n6cone}. However, the progress thus far has revealed only limited insight into structural patterns or physical principles, which undergird these constraints \cite{mmipt, her, arrangement, withxi, kbasis, superbalance, Akers:2021lms, withsirui, Fadel:2021urx, Hernandez-Cuenca:2022pst, He:2022bmi, gapholononholo}.

\smallskip
\textit{In this work---}
we formulate and prove two new infinite families of holographic entropy inequalities, one indexed by a pair of odd integers and another indexed by one arbitrary integer. These inequalities prove a prior conjecture that was motivated by unitary models of black hole evaporation, which identified a structural pattern in the holographic entropy cone \cite{withsirui}. Perhaps more importantly, the inequalities presented in this paper enjoy striking connections with many concepts from gravitational and condensed matter physics. Especially subregion duality \cite{subregion1, subregion2} and entanglement wedge nesting \cite{ewnref} are essential for formulating, proving, and understanding the inequalities. As we explain below, the inequalities appear to reflect a previously unknown topological consequence of entanglement wedge nesting.

\smallskip
\textit{Setup and notation.---} We consider a composite system with $m+n$ constituents, which are called $A_i$ ($1 \leq i \leq m$) and $B_j$ ($1 \leq j \leq n$). The indices on $A_i$ (resp. $B_j$) are understood modulo $m$ (resp. modulo $n$), for example $i = m+1 \equiv 1$. We assume that this $(m+n)$-partite system is in a pure state; this is equivalent to studying arbitrary states with $m+n-1$ constituents. 

Our inequalities are expressed in terms of unions of consecutive $A$- and $B$-type regions:
\begin{equation}
A_i^{(k)} = A_i A_{i+1} \ldots A_{i+k-1} 
\label{kterms}
\end{equation}
and likewise for $B_j^{(l)}$. When $m$ and $n$ are both odd, it is useful to shorthand a special case of this notation, which involves largest consecutive minorities and smallest consecutive majorities of $A$- and $B$-regions:
\begin{equation}
A_i^\pm \equiv A_i^{((m\pm 1)/2)} 
\qquad {\rm and} \qquad 
B_j^\pm \equiv B_j^{((n\pm 1)/2)}
\end{equation}
Throughout the paper, we write the inequalities in the form \mbox{`${\rm LHS} \geq {\rm RHS}$'} with only positive coefficients. 
\smallskip

\textit{New inequalities.---} We prove two new infinite families of holographic inequalities:

\begin{itemize}
\item 
{\bf Toric inequalities} are defined for $m$ and $n$, which are both odd. They take the following form:
\begin{equation}
\sum_{i=1}^m \sum_{j=1}^n S_{A_i^+ B_j^-}
\geq 
\sum_{i=1}^m \sum_{j=1}^n S_{A_i^- B_j^-} \,+ S_{A_1A_2\ldots A_m}
\label{toricineqs}
\end{equation}
The toric character of (\ref{toricineqs}) is related to the symmetry $\mathbb{Z}_m \times \mathbb{Z}_n$, which rotates the $A$- and $B$-regions. (The full symmetry group of (\ref{toricineqs}) is in fact $D_m \times D_n$.) 

This family subsumes the dihedral inequalities \cite{hec}
\begin{equation}
\sum_{i=1}^m S_{A_i^+}  \geq \sum_{i=1}^m S_{A_i^-} + S_{A_1 A_2 \ldots A_m}
\label{cyclic}
\end{equation}
by setting $n=1$ because $B_j^{(0)} = \emptyset$. Inequalities~(\ref{toricineqs}) also subsume two other previously known inequalities \cite{cuenca, withyunfei}; see Supplemental Material \cite{sm}.

\item
{\bf $\mathbb{RP}^2$ inequalities} are indexed by $m = n$:
\begin{align} 
\frac{1}{2} 
& \sum_{i,j=1}^{m}
\left(S_{A_i^{(j)} B_{i+j-1}^{(m-j)}} + S_{A_i^{(j)} B_{i+j}^{(m-j)}} \right) \nonumber \\
\geq &  
\sum_{i,j=1}^{m} S_{A_i^{(j-1)} B_{i+j-1}^{(m-j)}}
\, + S_{A_1A_2\ldots A_m}
\label{rp2ineqs}
\end{align}
We explain momentarily how (\ref{rp2ineqs}) relates to the projective plane. These inequalities are invariant under $D_{2m}$, which acts on the regions $B_1, A_1, B_2, A_2 \ldots B_m, A_m$, in this order, like it does on vertices of a regular $(2m)$-gon.

Family~(\ref{rp2ineqs}) includes monogamy of mutual information~(\ref{mmi}) and one other previously known inequality \cite{cuenca} as special cases ($m=2,3$).
\end{itemize}
We sketch a proof of (\ref{toricineqs}) and (\ref{rp2ineqs}) in the main text and fill in details in Supplemental Material \cite{sm}. 

\begin{figure}[t]
    \centering
    \scalebox{1}[1]{\includegraphics[width=0.8\linewidth]{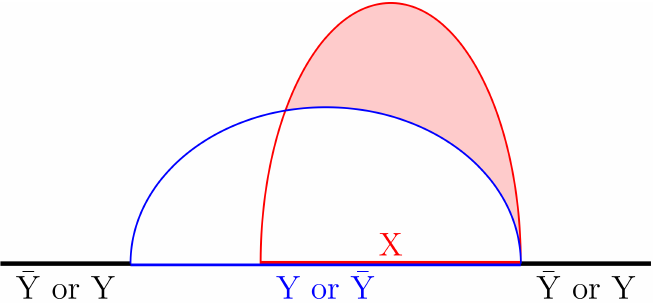}}
    \caption{Intersecting minimal cuts, which are forbidden by the entanglement wedge nesting theorem.}
    \label{fig:ewn}
\end{figure}

\smallskip
\textit{Entanglement wedge nesting (EWN).---} 
Schematically, each holographic inequality states the following: For any set of minimal cuts that realize the LHS terms, there exist some cuts for the RHS terms with a smaller or equal combined area. With this characterization of the problem, we should expect that facts and theorems about minimal cuts are likely to inform the structure and proofs of holographic entropy inequalities. This paper exploits one such fact, known as the entanglement wedge nesting (EWN) theorem \cite{ewnref}.

The theorem says that if two composite regions $X, Y$ are in either of these two relationships
\begin{equation}
X \subset Y \quad {\rm or} \quad X \subset \overline{Y}
\label{nestingcond}
\end{equation}
then the minimal cuts for $S_X$ and $S_Y$ can meet but cannot intersect; see Figure~\ref{fig:ewn}. In the AdS/CFT correspondence, subregion duality \cite{subregion1, subregion2} states that the largest bulk region reconstructible with access to boundary subsystem $X$---the entanglement wedge of $X$---is enclosed by $X$ and the minimal cut for $X$. This statement logically requires the EWN theorem or else extra access to $Y \setminus X$ would limit one's ability to reconstruct the bulk. 

As we explain below, inequalities~(\ref{toricineqs}) and (\ref{rp2ineqs})---and the crux of their proof---concern structural constraints on minimal cuts, which are induced by entanglement wedge nesting and which arise for topological reasons.

\begin{figure}
		\centering
		\includegraphics[width=1\linewidth]{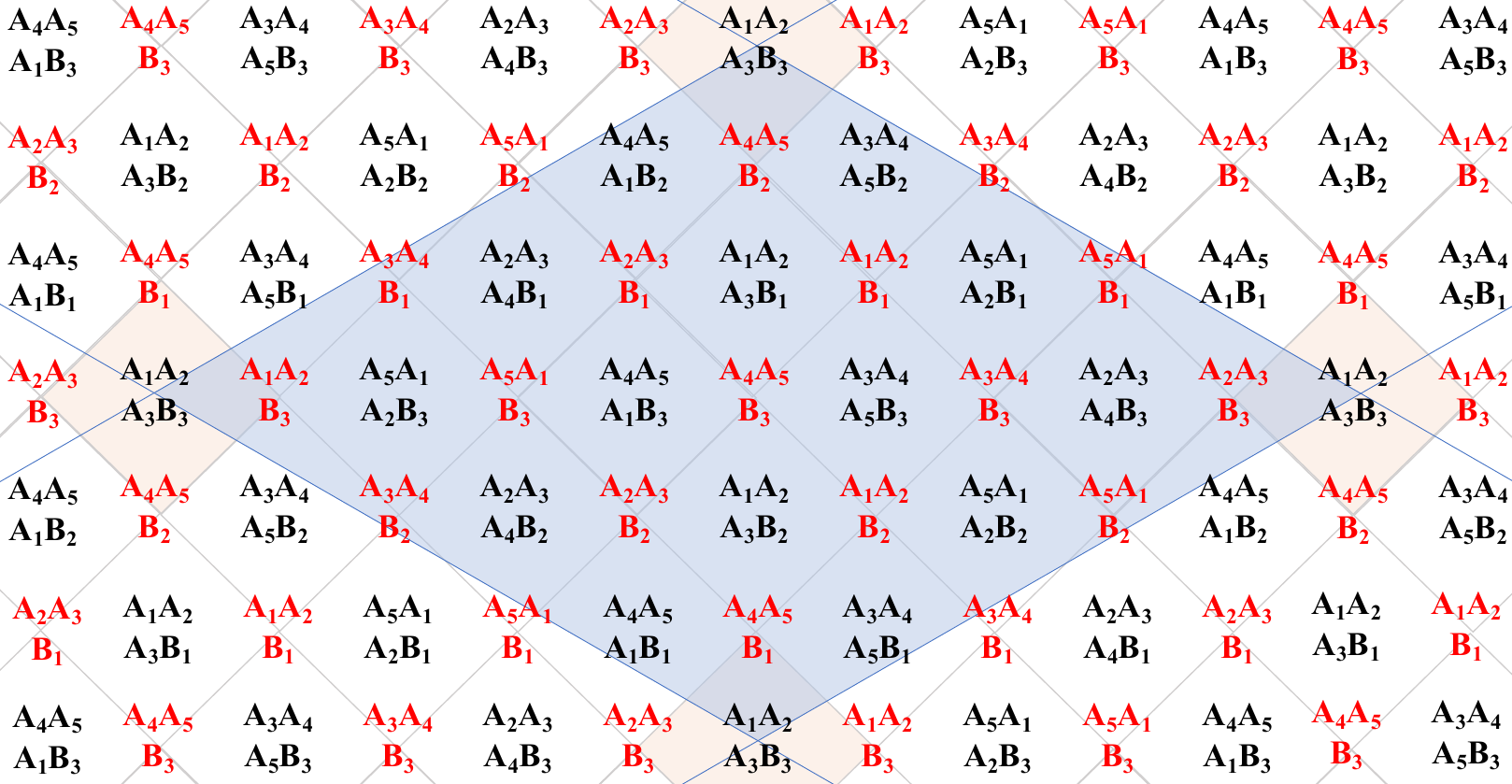} \\
		\vspace*{1mm}
		\includegraphics[width=1\linewidth]{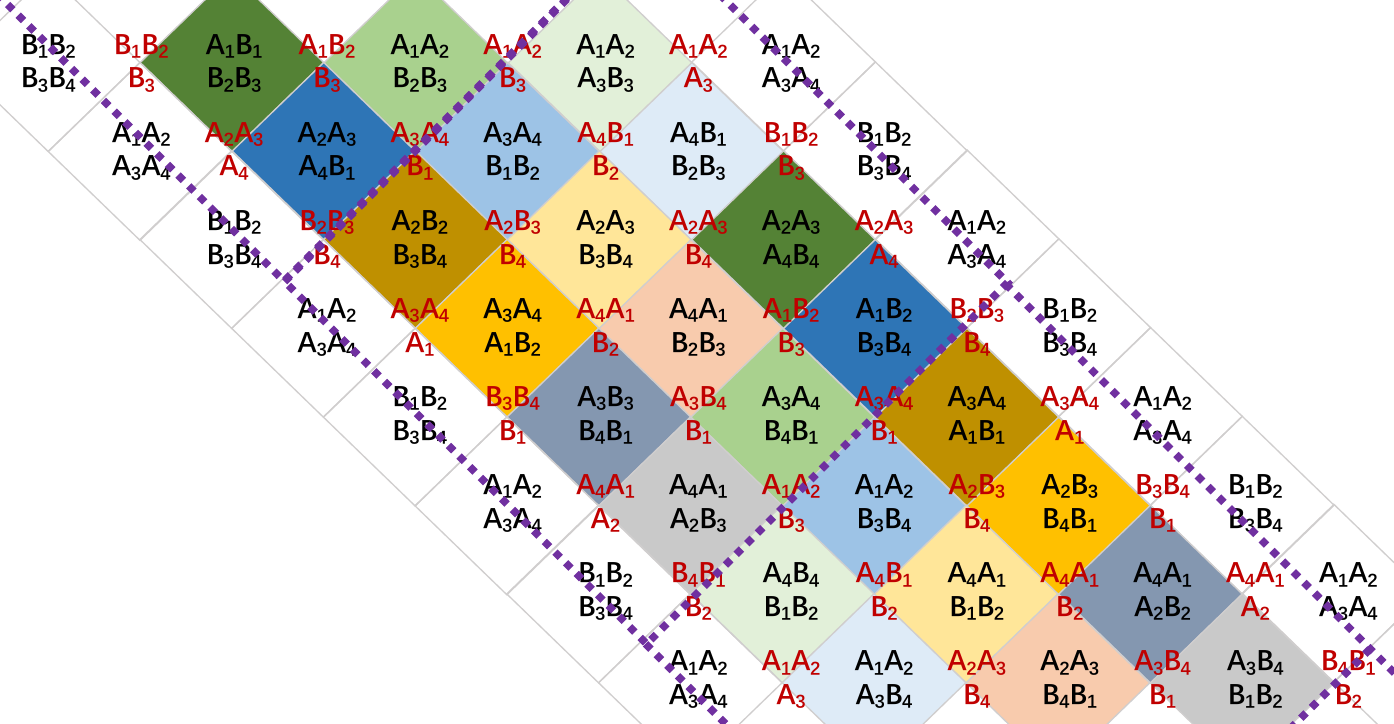}
	\caption{Graphical representations of a toric ($(m,n) = (5,3)$; upper panel) and an $\mathbb{RP}^2$ inequality ($m=4$; lower panel). Diamonds of same color are equivalent. We highlight fundamental domains of periodic identifications.} 
\label{fig:examples}
\end{figure}

\smallskip
\textit{Geometric organization.---} We propose a graphical way to organize terms of holographic inequalities, which is based on entanglement wedge nesting. We represent LHS terms as facets and RHS terms as vertices of a graph. Facets are defined with respect to an embedding on a two-dimensional surface, which we discuss momentarily. The organizing rule for the graph is that a vertex (RHS term) is incident to a facet (a LHS term) if and only if they satisfy a nesting relationship~(\ref{nestingcond}). 

A priori, it is not guaranteed that applying this rule to a holographic inequality will produce a sensible graph, which can be embedded on a two-dimensional manifold. Yet it turns out to work perfectly for inequalities~(\ref{toricineqs}) and (\ref{rp2ineqs}), with illuminating consequences.

The result of applying this graphical rule to inequalities~(\ref{toricineqs}) is shown in the upper panel of Figure~\ref{fig:examples}. It is a rhombus with diagonals spanning $m$ and $n$ squares and with opposite sides identified---a tessellation of a torus. The special term $S_{A_1 A_2 \ldots A_m}$ is left out of the graph because it is not in a nesting relationship with other terms. 

For inequalities~(\ref{rp2ineqs}), the result is shown in the lower panel of Figure~\ref{fig:examples}. A priori, the special RHS term $S_{A_1 A_2 \ldots A_m}$ could be canceled out because the LHS contains $m$ copies of the same. Taking inspiration from the toric inequalities, we do not cancel it but leave it out of the graph. This produces an $m \times m$ array of squares with one reflected periodic identification---a tessellation of the M{\"o}bius strip. The identification involves complementary regions with equal entropies, for example:
\begin{equation}
S_{A_i^{(j)} B_{i+j}^{(m-j)}} = S_{A_{i+j}^{(m-j)} B_{i}^{(j)}}
\label{rp2complements}
\end{equation}
The boundary of this M{\"o}bius strip consists of $m$ identical terms $S_{A_1 A_2 \ldots A_m} = S_{B_1 B_2 \ldots B_m}$. If we treat them all as one facet, we effectively glue the M{\"o}bius strip to a disk, which produces the projective plane $\mathbb{RP}^2$. 

\smallskip
\textit{Sketch of proof of inequalities.---} 
Reference~\cite{hec} formulated a general procedure for proving holographic entropy inequalities. (In the AdS$_{d+1}$/CFT$_d$ correspondence in $d > 2$, the procedure assumes that the quantum state in question is time reversal-symmetric \cite{hrt, withxi}.) In essence, we divide minimal cuts for LHS terms into segments, which are then reassembled into cuts for RHS terms. 

To sketch the proof of inequalities~(\ref{toricineqs}) and (\ref{rp2ineqs}), we need to briefly review the mechanics of this procedure. Each segment of a LHS minimal cut is associated to a pair of bit strings $x, x' \in \{0,1\}^l$, where $l$ is the number of LHS terms in the inequality, accounting for multiplicity. The bit strings $x, x'$ characterize the bulk regions on both sides of the LHS cut, with a 1 (respectively 0) in the $k^{\rm th}$ bit when the region is inside (outside) the entanglement wedge of the $k^{\rm th}$ LHS term. It follows that $|x-x'|_1$ counts how many times the given cut segment contributes on the left hand side of the inequality. We now construct RHS cuts. To do so, a function $f: \{0,1\}^l \to \{0,1\}^r$ is introduced, where $r$ counts RHS terms with multiplicity. The interpretation of bits in $f(x) \in \{0,1\}^r$ mirrors that of $x \in \{0,1\}^l$: the $k^{\rm th}$ bit of $f(x)$ tells us that the bulk region labeled $x$ is outside/within (0/1) the cut for the $k^{\rm th}$ RHS term. Therefore, $|f(x)-f(x')|_1$ is the multiplicity with which the constructed RHS cuts use the LHS cut segment that divides $x$ from $x'$. To prove the inequality, one must find an $f$ such that
\begin{equation}
|x-x'|_1 \geq |f(x) - f(x')|_1 \qquad \textrm{for all}~~x, x' \in \{0, 1\}^l,
\end{equation}
i.e. $f(x)$ must be a \emph{contraction}. In addition, $f(x)$ must satisfy boundary conditions, which enforce the homology condition for the resulting RHS cuts. The boundary conditions are discussed in Supplemental Material \cite{sm}.  

\begin{figure}[t]
    \centering
    \raisebox{35mm}{
    $\begin{array}{c}
    \includegraphics[width=0.2\linewidth]{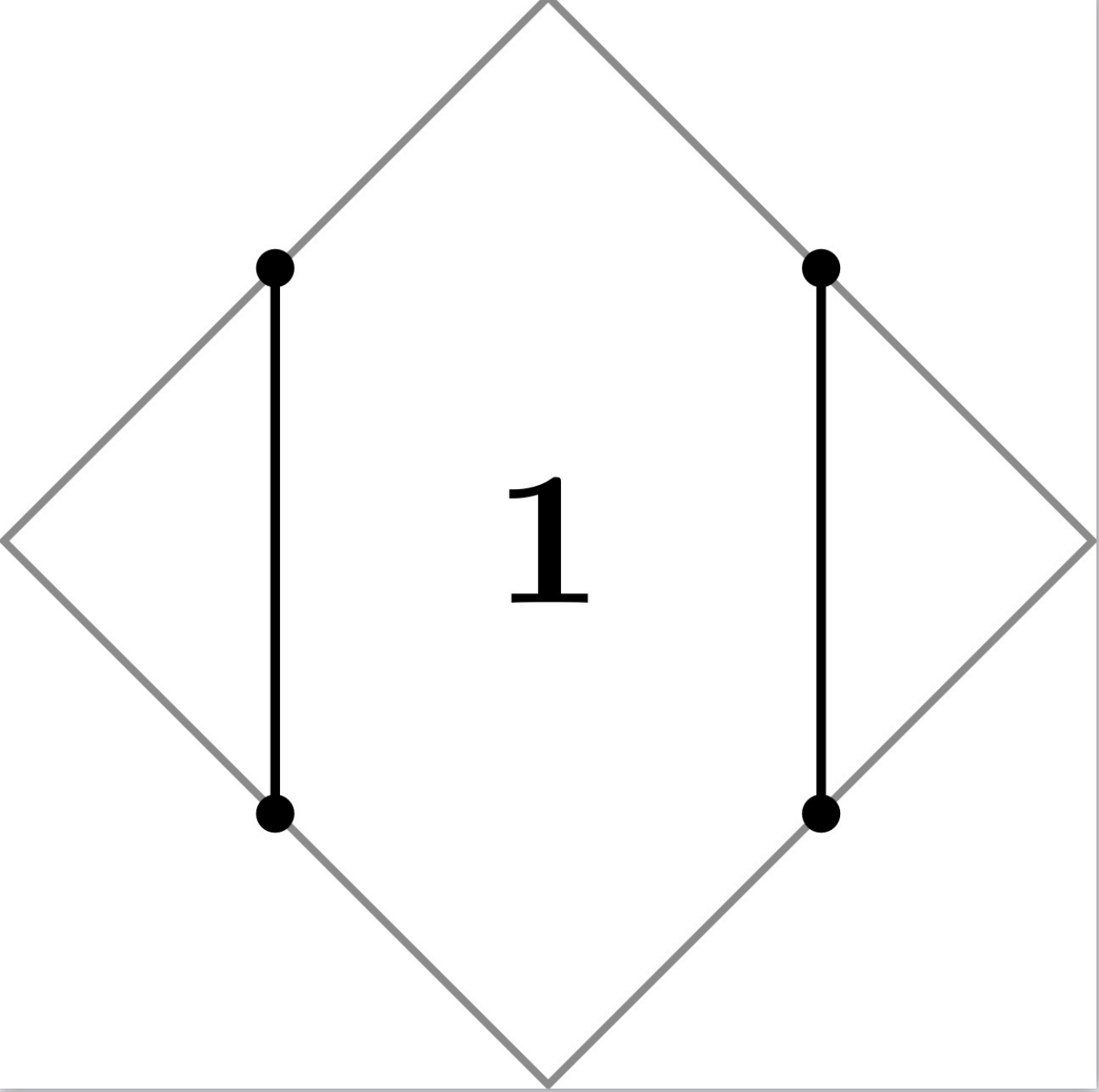} \\
    \includegraphics[width=0.2\linewidth]{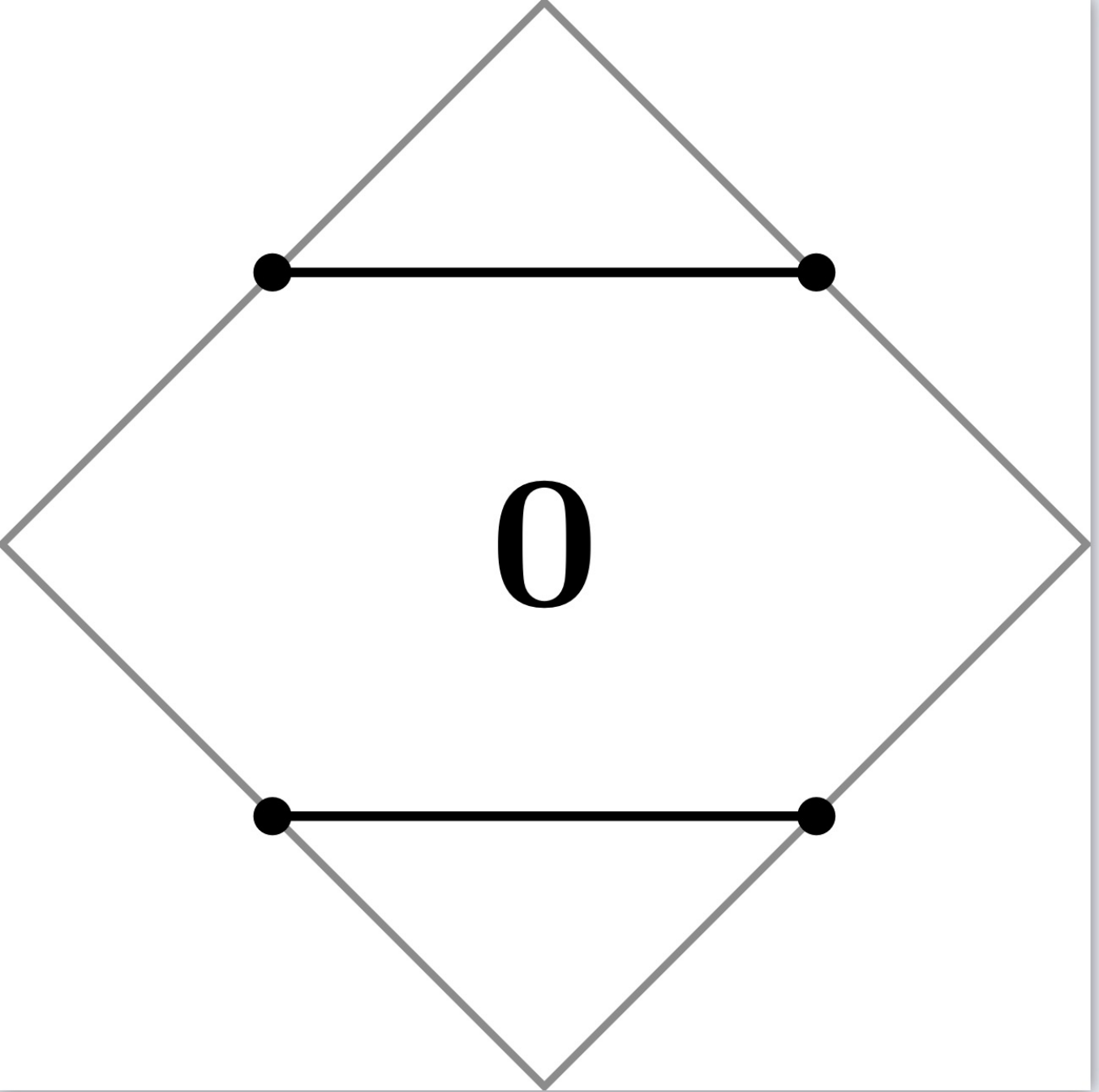}
    \end{array}$}\qquad
    \includegraphics[width=0.65\linewidth]{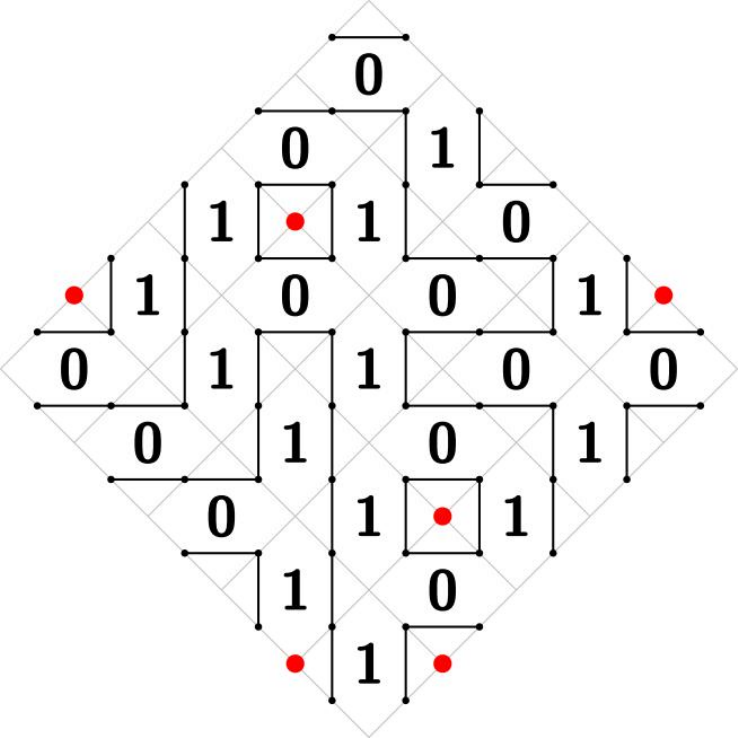}
    \caption{Rules for drawing graphs $\Gamma(x)$ and one such graph, which arises in proving the $(5,5)$ toric inequality. Tiling vertices, which are mapped to 1 under $f(x)$, are highlighted.}
    \label{fig:gamma}
\end{figure}

The contraction that proves inequalities~(\ref{toricineqs}) is defined as follows. (The proof of~(\ref{rp2ineqs}) is similar in spirit; see Supplemental Material.) Since our graphical scheme represents LHS terms as facets in a square tiling, every $x \in \{0,1\}^l$ is an assignment of 0's and 1's to squares. The contraction $f(x) \in \{0,1\}^r$ assigns 0 or 1 to every vertex in the tiling and to the special term $S_{A_1 A_2 \ldots A_m}$ not present in the tiling. To express $f(x)$, define a graph $\Gamma(x)$ by drawing a horizontal/vertical pair of line segments on every square that carries a 0/1, as shown in Figure~\ref{fig:gamma}. Because every node of $\Gamma(x)$---that is, every edge in the tiling---is connected to two other nodes, $\Gamma(x)$ is composed of nonintersecting loops. Therefore, $\Gamma(x)$ partitions the torus into components. Map $f$ assigns 1 to one special vertex (say, bottommost and rightmost) in every component, which does not wrap a nontrivial cycle of the torus. To all other vertices $f$ assigns 0. Finally, the value of $f$ in the last bit (on the special RHS term $S_{A_1 A_2 \ldots A_m}$) is set so that $|x|_1 \equiv |f(x)|_1$~(mod 2).

For details, and to confirm that $f$ is indeed a contraction, see Supplemental Material. Here we wish to convey that the spatial organization of terms as facets and vertices on a manifold is indispensable for proving the inequalities. The logic of the proof further suggests that we should view inequalities (\ref{toricineqs}) and (\ref{rp2ineqs}) as topological statements, which rely on entanglement wedge nesting. Considering the close relationship between nesting and erasure correction \cite{errorcorr, rtfromcorr}, the inequalities might be most usefully understood as topological constraints on holographic error correcting codes. Leaving this tentative thought for future work, let us take a more direct look at the meaning of the inequalities. 

\smallskip
\textit{Inequalities versus strong subadditivity.---}Consider one edge of the tiling in Figure~\ref{fig:examples}. Let the two facets incident to the edge be $S_X$ and $S_{\overline{V}}$. An inspection of the inequalities reveals that the vertices incident to the edge are $S_{\overline{X \cup V}}$ and $S_{X \cap V}$. For example, in inequalities~(\ref{toricineqs}), the edge that separates the facets with
\begin{equation}
X = A_i^+ B_{j}^- \quad {\rm and} \quad \overline{V} = A_{i+(m-1)/2}^+ B_{j+(n+1)/2}^- 
\label{defxv}
\end{equation}
has endpoints at 
\begin{equation*}
\overline{X \cup V} = A_{i+(m+1)/2}^- B_{j+(n+1)/2}^- \quad {\rm and} \quad X \cap V = A_i^- B_{j}^-.
\end{equation*}
The same observation holds in the graph of (\ref{rp2ineqs}). 

If we now collect the four terms that are associated with one edge, we obtain a nonnegative \cite{ssaref} quantity called conditional mutual information:
\begin{equation}
{\rm CMI}_{\rm edge} := S_X + S_V - S_{X \cap V} - S_{X \cup V}
\label{edgessa}
\end{equation}
We have used the fact that we are working with pure states so complementary regions have equal entropies. 

After adding up contributions from various edges, and correcting for a double-counting of terms, we discover that inequalities~(\ref{toricineqs}, \ref{rp2ineqs}) can be written in the simple form:
\begin{equation}
\tfrac{1}{2} \sideset{}{'}\sum {\rm CMI}_{\rm edges} \geq S_{A_1 A_2 \ldots A_m}
\label{ineqcmi}
\end{equation}
In toric inequalities, the $\sum'$ runs over either set of parallel edges. In the $\mathbb{RP}^2$ inequalities we sum over edges, which run parallel to the boundary of the M{\"o}bius strip in Figure~\ref{fig:examples}. Rewriting~(\ref{ineqcmi}) presents the inequalities as a collective improvement on strong subadditivity.

\smallskip
\textit{Continuum limit.---} The regular structure of the inequalities allows us to take a continuum limit $m,n \to \infty$. The idea is to keep the union of all $A$- and $B$-type regions fixed, but subdivide it into a growing number of $A_i$s and $B_j$s. Due to the cyclic symmetries that rotate the $A$- and $B$-regions, it is easiest to visualize a special case: an entangled state of two 1+1-dimensional holographic conformal field theories (CFT$_2$) living on circles, which are subdivided into intervals $A_i$ and $B_j$. Holographically, this setup describes a 2+1-dimensional, two-sided black hole with entropy $S_{\rm BH} = S_{A_1 A_2 \ldots A_m}$.

In these settings, studying the continuum limit of holographic inequalities has a useful precedent. Taking $m \to \infty$ in subfamily~(\ref{cyclic}) of the toric inequalities gives:
\begin{equation}
S_{\rm diff} = \oint dv \frac{\partial S(u,v)}{\partial v} \Big|_{u = u(v)} 
\geq S_{A_1 A_2 \ldots A_m} = S_{\rm BH},
\label{cycliclimit}
\end{equation}
where $S(u,v)$ is the entanglement entropy of interval $(u,v)$. In expression~(\ref{cycliclimit}), we traded the discrete label $i$ in $A_i^+$ for a continuous variable $v$ using $A_i^+ := (u(v),v)$, so that the function $u(v)$ implicitly encodes the sizes of intervals $A_i^+$. Quantity~$S_{\rm diff}$, called differential entropy \cite{diffent, hholes}, computes the length of the bulk curve whose tangents subtend boundary intervals $(u(v),v)$. Inequality~(\ref{cycliclimit}) is manifestly true because such a curve necessarily wraps around the black hole horizon. 

For a similar limit of inequalities~(\ref{toricineqs}), let $A_i^+ = (u_A, v_A(u_A))$ and $B_j^- = (u_B(v_B), v_B)$. Observing that in the continuum limit quantity ${\rm CMI}_{\rm edge}$ becomes a second partial derivative of entropy, we find:
\begin{equation}
- \frac{1}{2} \oiint du_A dv_B 
\frac{\partial^2 S(u_A, v_A; u_B, v_B)}{\partial u_A\, \partial v_B} 
\geq S_{\rm BH}
\label{toriclimit}
\end{equation}
The integral in (\ref{toriclimit}) admits many bulk interpretations, depending on the phases of $S(u_A, v_A; u_B, v_B)$. Its behaviors---as well as the continuum limit of inequalities~(\ref{rp2ineqs})---will be studied in a separate publication~\cite{withdachen}. 

To give a flavor of the bulk interpretation of~(\ref{toriclimit}), we consider one illustrative case. 
Assume that the Ryu-Takayanagi surfaces for $A_i^\pm B_j^-$ do not wrap or cross the horizon; this can be easily arranged by choosing a shockwave geometry \cite{shockwaves} with a large horizon. Then the terms in (\ref{toricineqs}) can only be in one of two phases:
\begin{equation*}
S_{A_i^\pm B_j^-} = 
\big(S_{A_i^\pm} + S_{B_j^-}\big)~~{\rm or}~~\big(S_{A_{i+(m\pm 1)/2}^\mp} + S_{B_{j+(n-1)/2}^+}\big)
\label{termcases}
\end{equation*} 
Suppose $B_n$ is much larger than all the other regions, so that the phase of $S_{A_i^\pm B_j^-}$ is determined solely by whether or not $B_n \subset B_j^-$. Substituting in eqs.~(\ref{defxv}, \ref{edgessa}), we see that the only nonvanishing terms in (\ref{ineqcmi}) arise from $X = A_i^+ B_{(n+1)/2}^-$ and $\overline{V} = A_{i+(m-1)/2}^+ B_1^-$ and evaluate to:
\begin{align}
{\rm CMI}_{\rm edge} 
= \, & 
\big(S_{A_i^+} - S_{A_i^-}\big) + \big(S_{A_{i+(m-1)/2}^+} - S_{A_{i+(m+1)/2}^-}\big) \nonumber \\
\xrightarrow{\rm continuum}\,\,\, & 
dv_A \frac{\partial S(u_A,v_A)}{\partial v_A} - du_A \frac{\partial S(u_A,v_A)} {\partial u_A}
\label{continuumintegrand}
\end{align}
Replacing $\tfrac{1}{2} \sum' \to \tfrac{1}{2}\oint$ and integrating by parts, we again find (\ref{cycliclimit}). Thus, toric inequalities also reproduce the geometric fact `differential entropy $\geq$ black hole entropy.' 

\smallskip
\textit{Significance for the holographic entropy cone.---} In our search for order among holographic inequalities, two of us previously conjectured a different pattern \cite{withsirui} (see also \cite{Fadel:2021urx}). Take any valid inequality and apply to it the permutation group acting on region labels. Adding up all permutation images gives a weaker inequality, which involves only averages of $p$-partite entropies, denoted $S^p$. For example, averaging monogamy~(\ref{mmi}) in this way gives:
\begin{align}
(S_{AB} + S_{BC} + S_{CA}) & \geq (S_A + S_B + S_C) + S_{ABC} \nonumber \\
\rightarrow\quad 3S^2 \phantom{S_{C} + S)} & \geq \phantom{ ((S_{A} +} 3 S^1 \phantom{+ S_C))} + S^3
\end{align}
We conjectured that for every $p$ there exists a valid holographic inequality on $2p-1$ or more regions which, after averaging, gives:
\begin{equation}
2 S^p / p \geq S^{p-1} / (p-1) + S^{p+1}/(p+1)
\label{avineqs}
\end{equation}
If valid, inequalities~(\ref{avineqs}) cannot be improved. That is, every valid holographic inequality averages to a convex combination of (\ref{avineqs}). This is because so-called extreme rays---values of $S^p$ that simultaneously saturate all but one inequality~(\ref{avineqs})---have known realizations as tensor networks and holographic geometries, so any improvement over (\ref{avineqs}) is necessarily wrong. These extreme rays correspond rigorously to stages of evaporation of an old black hole---a fact that motivated our conjecture. 

Each family presented in this paper independently proves the conjecture. Indeed, toric inequalities with $m = n$ and $m = n+2$ average to (\ref{avineqs}) with $p = (m+n)/2$ while the $\mathbb{RP}^2$ inequalities average to (\ref{avineqs}) with $p=m$. 

At present, we know 375 holographic inequalities, which are not part of families (\ref{toricineqs}, \ref{rp2ineqs}). It will be interesting to see whether they too are associated with two-dimensional manifolds under our EWN-based graphical scheme. We have not yet checked this because 373 of them were announced only days ago \cite{n6cone}. Encouragingly, the answer is affirmative for one of the `older' inequalities: it defines a polytope with the topology of the $g=2$ Riemann surface; see Supplemental Material \cite{sm}.

The final comment asks whether inequalities~(\ref{toricineqs}, \ref{rp2ineqs}) might be improved. We expect that they are maximally tight but proving so appears to be challenging. Note, however, that every individually studied, previously known instance of these inequalities turned out to be maximally tight; see Supplemental Material.

\begin{acknowledgements}
\noindent
We thank for useful discussions Bowen Chen, Sergio Hern\'andez-Cuenca, Ricardo Esp{\'i}ndola, Matthew Headrick, Veronika Hubeny, Yunfei Wang, Dachen Zhang as well as participants of the `Quantum Information, Quantum Matter and Quantum Gravity' workshop (YITP-T-23-01) held at YITP, Kyoto University (2023), where this work was completed. YW acknowledges support from the Shuimu Tsinghua Scholar Program of Tsinghua University. This research was supported by an NSFC grant number 12042505, a BJNSF grant under the Gao Cengci Rencai Zizhu program, and a Dushi Zhuanxiang Fellowship. BC dedicates this paper to Bayu Mi{\l}osz Czech in celebration of his seventh birthday.
\end{acknowledgements}

\medskip

\cleardoublepage
\section*{Supplemental Material}

\subsection{Relations to previously known inequalities}
\label{comparison}

For the reader's convenience, we copy from the main text the toric inequalities
\begin{equation}
\sum_{i=1}^m \sum_{j=1}^n S_{A_i^+ B_j^-}
\geq 
\sum_{i=1}^m \sum_{j=1}^n S_{A_i^- B_j^-} \,+ S_{A_1A_2\ldots A_m}
\label{app:toricineqs}
\end{equation}
and the $\mathbb{RP}^2$ inequalities:
\begin{align} 
\frac{1}{2} 
& \sum_{j,i=1}^{m}
\left(S_{A_i^{(j)} B_{i+j-1}^{(m-j)}}  + S_{A_i^{(j)} B_{i+j}^{(m-j)}}\right) \nonumber \\
\geq &  
\sum_{j,i=1}^{m} S_{A_i^{(j-1)} B_{i+j-1}^{(m-j)}}
\, + S_{A_1A_2\ldots A_m}
\label{app:rp2ineqs}
\end{align}
In (\ref{app:rp2ineqs}), the LHS double counts identical terms (see equation~(\ref{rp2complements}) from the main text), then corrects this with a factor of $1/2$. When writing out inequalities explicitly, it is practical to break symmetry and choose one representative in each doublet of identical terms. This yields the following rewriting of (\ref{app:rp2ineqs}):
\begin{align}
& (m-1) S_{A_1^{(m)}} + \!\!
\sum_{j=1}^{\myfloor{m/2}}\!\! \sum_{i=1}^{~i_{\rm max}(j)}
\!\!\left(S_{A_i^{(j)} B_{i+j-1}^{(m-j)}}  + S_{A_i^{(j)} B_{i+j}^{(m-j)}} \right)
\nonumber \\
& \quad \geq 
\sum_{j,i=1}^m S_{A_i^{(j-1)} B_{i+j-1}^{(m-j)}}
\label{rp2asymm}
\end{align}
Here $i_{\rm max}(j) = m$ unless $j = m/2$ with $m$ even, in which case $i_{\rm max}(j) = m/2$. This cooky summation limit occurs because for even $m$ the terms at $j = m/2$ get identified with one another under the $\mathbb{Z}_2$ identification, which produces the M{\"o}bius strip. 

\subsubsection{Lowest toric inequalities}

{\bf Dihedral inequalities}~~Substituting $(m,n) = (1,1)$ in (\ref{app:toricineqs}) yields a true if prosaic statement:
\begin{equation}
S_{A_1} \geq S_{A_1}
\end{equation}

The $(m,n) = (3,1)$ toric inequality is the monogamy of mutual information:
\begin{equation}
S_{A_1 A_2} + S_{A_2 A_3} + S_{A_3 A_1} \geq S_{A_1} + S_{A_2}  + S_{A_3} + S_{A_1 A_2 A_3}
\label{app:mmi}
\end{equation}

Ineq.~(\ref{app:mmi}), as well as all the higher $(m,1)$ toric inequalities, are the dihedral inequalities \cite{hec}, which were written in (\ref{cyclic}) in the main text.

\medskip
{\bf The $(3,3)$ toric inequality}~~was listed as number~5 in \cite{cuenca}. The same reference proved it to be maximally tight.
\begin{align} 
   S_ {ABC} \,+\, & S_ {ACD}  + S_ {ADB} \nonumber \\
+\, S_ {EBC} \,+\, & S_ {ECD} + S_ {EDB} \nonumber \\      
+\, S_ {EAD} \,+\, & S_ {EAB}  + S_ {EAC} \nonumber \\
& \geq \tag{$i_5$} \label{i5} \\
     S_ {AB} \,+\, & S_ {AC} + S_ {AD} \nonumber \\
+\, S_ {EB} \,+\, & S_ {EC}   +   S_ {ED} \nonumber \\
   +\,   S_ {EACD}  \,+\,  & S_ {EADB}   +   S_ {EABC}   \nonumber \\
   \,+\, & S_ {BCD} \nonumber
\end{align}
To highlight the symmetry, we introduce a purifier $O$ and rewrite some terms using complementary regions:
\begin{align} 
   S_ {ABC} \,+\, & S_ {ACD}  + S_ {ADB} \nonumber \\
+\, S_ {EBC} \,+\, & S_ {ECD} + S_ {EDB} \nonumber \\      
+\, S_ {OBC} \,+\, & S_ {OCD}  + S_ {ODB} \nonumber \\
& \geq \tag{$i_5'$} \label{i5p} \\
     S_ {AB} \,+\, & S_ {AC} + S_ {AD} \nonumber \\
+\, S_ {EB} \,+\, & S_ {EC}   +   S_ {ED} \nonumber \\
   +\,   S_ {OB}  \,+\,  & S_ {OC}   +   S_ {OD}   \nonumber \\
   \,+\, & S_ {AEO} \nonumber
\end{align}
The 3x3 arrays of terms realize two dihedral symmetries $D_3 = S_3$, which act independently on their horizontal and vertical directions. Comparing (\ref{i5p}) with (\ref{app:toricineqs}), we see that the two expressions match with $(A_1A_2A_3)\, (B_1 B_2 B_3) \leftrightarrow (AEO)(BCD)$. 

\medskip
{\bf The $(m,n) = (5,3)$ inequality}~~was announced in \cite{withyunfei}, who also proved that it is maximally tight.
\begin{align}
     S_ {ADEF} + S_ {BCGE} \,+\, & S_ {BCDE} + S_ {BCDF} + S_ {AGDE}  \nonumber \\
+\, S_ {BDEF} + S_ {ACGE} \,+\, & S_ {ACDE} + S_ {ACDF} + S_ {BGDE}  \nonumber \\
+\, S_ {ADEF} + S_ {ABGE} \,+\, & S_ {ABDE} + S_ {ABDF} + S_ {CGDE}  \nonumber \\
& \geq  \\
     S_ {ADE} + S_ {ADF} \,+\, & S_ {BCDEG} + S_ {BCDEF} + S_{AEG} \nonumber  \\
+\, S_ {BDE} + S_ {BDF} \,+\, & S_ {ACDEG} + S_ {ACDEF} + S_{BEG} \nonumber  \\  
+\, S_ {CDE} + S_ {CDF} \,+\, & S_ {ABDEG} + S_ {ABDEF} + S_{BEG} \nonumber  \\
\,+\, & S_ {ABC}  \nonumber
\end{align}
We introduce a purifier $O$, substitute some terms for their complements, and neglect lexicographic order in certain indices to finally obtain:
\begin{align}
     S_ {AEDF} + S_ {ADFO} \,+\, & S_ {AFOG} + S_ {AOGD} + S_ {AGED}  \nonumber \\
+\, S_ {BEDF} + S_ {BDFO} \,+\, & S_ {BFOG} + S_ {BOGD} + S_ {BGED}  \nonumber \\
+\, S_ {CEDF} + S_ {CDFO} \,+\, & S_ {CFOG} + S_ {COGD} + S_ {CGED}  \nonumber \\
& \geq  \\
     S_ {AED} + S_ {ADF} \,+\, & S_ {AFO} + S_ {AOG} + S_{AGE} \nonumber  \\
+\, S_ {BED} + S_ {BDF} \,+\, & S_ {BFO} + S_ {BOG} + S_{BGE} \nonumber \\  
+\, S_ {CED} + S_ {CDF} \,+\, & S_ {CFO} + S_ {COG} + S_{CGE}  \nonumber \\
\,+\, & S_ {EDFOG}  \nonumber
\end{align}
This matches the form of~(\ref{app:toricineqs}) under the replacement: 
\begin{equation*}
(EDFOG)\,(ABC) \leftrightarrow (A_1A_2A_3 A_4 A_5)\,(B_1B_2B_3)
\end{equation*}

\subsubsection{Lowest projective plane inequalities}
Substituting $m=1$ in (\ref{rp2asymm}) gives $0 \geq 0$.

\medskip
{\bf The $m=2$ inequality}~~written in the form~(\ref{rp2asymm}) is:
\begin{equation}
S_{A_1 A_2} + S_{A_1 B_1} + S_{A_1 B_2} \geq S_{B_1} + S_{B_2} + S_{A_1} + S_{A_2}
\end{equation}
Once again, this is the monogamy of mutual information.

\medskip
{\bf The $m=3$ inequality}~~was previously announced as number 7 in \cite{cuenca}, in the form:
\begin{align}
2S_{ABC} 
\,+\, & S_{BCE} + S_{ADE} \nonumber \\
\,+\, & S_{BDE} + S_{ACD} \nonumber \\
\,+\, & S_{ABD} + S_{ABE} \nonumber \\
& \quad\geq \tag{$i_7$} \label{i7} \\
S_{ABCE} \,+\, & S_{DE} + S_{ABCD} \nonumber \\
S_{AD} \,+\, & S_{BE} + S_{ABDE} \nonumber \\
S_{AB} \,+\, & S_{BC} + S_{AC} \nonumber
\end{align}
Ref.~\cite{cuenca} proved that (\ref{i7}) is maximally tight. It matches inequality~(\ref{rp2asymm})
\begin{align}
2S_{A_1 A_2 A_3} 
\,+\, & S_{A_1 B_1 B_2} + S_{A_1 B_2 B_3} \nonumber \\
\,+\, & S_{A_2 B_2 B_3} + S_{A_2 B_3 B_1} \nonumber \\
\,+\, & S_{A_3 B_3 B_1} + S_{A_3 B_1 B_2} \nonumber \\
& \quad\geq \\
S_{B_1 B_2} \,+\, & S_{B_2 B_3} + S_{B_3 B_1} \nonumber \\
S_{A_1 B_2} \,+\, & S_{A_2 B_3} + S_{A_3 B_1} \nonumber \\
S_{A_1 A_2} \,+\, & S_{A_2 A_3} + S_{A_3 A_1} \nonumber
\end{align}
under the identification
\begin{equation*}
(ABC)\,(ODE) \leftrightarrow (A_1 A_2 A_3)\,(B_1 B_2 B_3),
\end{equation*} 
where $O$ is the purifying region.

\begin{figure}[t]
\centering
\includegraphics[width=0.68\linewidth]{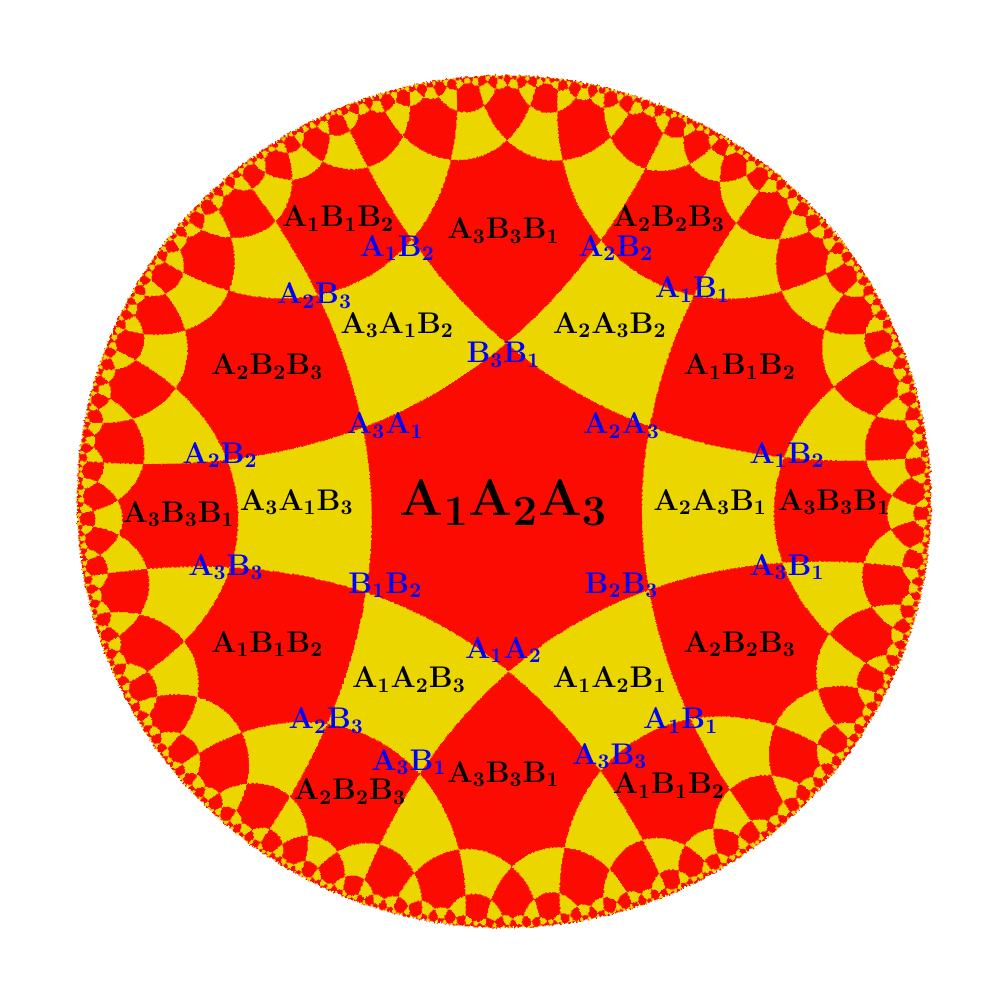}
\caption{Graphical arrangement of (\ref{i6}) displayed on the hyperbolic disk, which is its universal cover.}
 \label{fig:i6}
 \end{figure}

\subsubsection{A graphical representation of another inequality}
Here we apply our EWN-based graphical representation to one of the two holographic inequalities, which have been publicly known for more than a week and which are not covered by families (\ref{app:toricineqs}) and (\ref{app:rp2ineqs}). This inequality was listed as number 6 in \cite{cuenca}:
\begin{align} 
       3S_ {ACE} \,+\, & 3S_ {ABC} + 3S_ {ABD} \nonumber \\
        +\, S_ {BCE} \,+\, & S_ {ABE} + S_ {CDE} \nonumber \\
        +\, S_ {BCD} \,+\, & S_ {BDE} + S_ {ACD} \nonumber \\
        +\,& S_ {ADE} \nonumber \\
        & \ge    \tag{$i_6$} \label{i6} \\
        S_ {AD} \,+\, & S_ {AE} + S_ {DE}  \nonumber \\
        +\, 2S_ {ABCE} \,+\, & 2S_ {AB} + 2S_ {CE} \nonumber \\
        +\, 2S_ {BD} \,+\, & 2S_ {AC} + 2S_ {ABCD} \nonumber \\
        +\, S_ {ACDE} \,+\, & S_{BC} + S_ {ABDE} \nonumber  
   \end{align}
To highlight its structure, let us replace
\mbox{$(DAE)\,(OBC) \leftrightarrow (A_1A_2A_3)\,(B_1B_2B_3)$}
and trade certain terms for their complements:
\begin{align} 
        3S_ {A_1B_1B_2} \,+\, & 3S_ {A_2B_2B_3} + 3S_ {A_3B_3B_1} \nonumber \\
        +\, S_ {A_1A_2B_1} \,+\, & S_ {A_2A_3B_2} + S_ {A_3A_1B_3} \nonumber \\
        +\, S_ {A_1A_2B_3} \,+\, & S_ {A_2A_3B_1} + S_ {A_3A_1B_2} \nonumber \\
        +\,& S_ {A_1A_2A_3} \nonumber \\
        & \ge    \label{i6ours} \\
        S_ {A_1A_2} \,+\, & S_ {A_2A_3} + S_ {A_3A_1}  \nonumber \\
        +\, 2S_ {A_1B_1} \,+\, & 2S_ {A_2B_2} + 2S_ {A_3B_3}\nonumber \\
        +\,2S_ {A_1B_2} \,+\, & 2S_ {A_2B_3} + 2S_ {A_3B_1} \nonumber \\
        +\,S_ {B_1B_2} \,+\, & S_ {B_2B_3} + S_{B_3B_1} \nonumber  
   \end{align}

The graphical representation of this inequality is shown in Figure~\ref{fig:i6}. We draw it as a tessellation of the hyperbolic disk, modded out by a certain isometry subgroup. Experts on hyperbolic geometry will be able to read off directly from Figure~\ref{fig:i6} that the graph of inequality~(\ref{i6ours}) is a discretization of the $g=2$ Riemann surface.

We can establish the same fact in a more pedestrian way by counting vertices, faces and edges:
\begin{equation}
V = 12 \qquad {\rm and} \qquad F = 10 \qquad {\rm and} \qquad E = 24
\end{equation}
We find $\chi = V + F - E = -2 = 2 - 2g$ so this graph is embedded on the $g=2$ Riemann surface. The polytope defined by inequality~(\ref{i6ours}) is called cubohemioctahedron.

\medskip
{\bf Remarks}~~
Relative to inequalities~(\ref{app:toricineqs}) and (\ref{app:rp2ineqs}), one novelty in Figure~\ref{fig:i6} is the appearance of non-square faces. Another is that faces and edges, which are related by a symmetry of the cubohemioctahedron, nevertheless appear in (\ref{i6ours}) with different coefficients. Our graphical scheme knows only about nesting relationships among terms and neglects coefficients entirely.

We do not know how to generalize structure~(\ref{i6ours}) and Figure~\ref{fig:i6} to larger numbers of regions. 

\subsection{Proof of toric inequalities}
\label{ineqsproof}

\subsubsection{Generalities}
{\bf Contraction maps}~~This method of proof was developed in \cite{hec}. 
For each inequality to be proven, we must find a contraction map $f: \{0,1\}^l \to \{0,1\}^r$ such that 
\begin{equation}
|x-x'|_1 \geq |f(x) - f(x')|_1 \qquad \textrm{for all}~~x, x' \in \{0, 1\}^l
\end{equation}
Here $l$ (respectively $r$) counts terms on the left (right) hand side of the inequality, accounting for multiplicity. Bits in $x \in \{0,1\}^l$ and in $f(x) \in \{0,1\}^r$ are associated with specific terms in the inequality and, therefore, with facets (respectively vertices) in the tiling. The only exception is the last bit in $f(x)$, which is associated with the term $S_{A_1 A_2 \ldots A_m}$ on the RHS and which is not represented in the tiling.  

We note one lemma \cite{hec}, which simplifies the task of verifying that $f$ is a contraction:
\begin{align}
{\rm if}~|x-x'|_1 & \geq |f(x) - f(x')|_1 \quad \textrm{for all}~~|x-x'|_1 = 1 \nonumber \\
{\rm then}~|x-x'|_1 & \geq |f(x) - f(x')|_1 \quad \textrm{for all}~~x, x'
\label{simplification}
\end{align}
In other words, it suffices to check the contracting property on $x$'s, which differ in a single bit. 

\medskip
{\bf Boundary conditions}~~For each region $A_i$ and $B_j$, we define an indicator $x^{A_i}$ (respectively $x^{B_j}$) by setting the $k^{\rm th}$ bit of it to:
\begin{equation}
(x^{A_i})_k = \begin{cases} 
1 & {\rm if}~A_i \subset~k^{\rm th}~\textrm{LHS term} \\ 
0 & {\rm otherwise}
\end{cases}
\label{xindicator}
\end{equation}
Similarly, we define indicators $f^{A_i}$ and $f^{B_j}$ by setting their $k^{\rm th}$ bits to:
\begin{equation}
(f^{A_i})_k = \begin{cases} 
1 & {\rm if}~A_i \subset~k^{\rm th}~\textrm{RHS term} \\ 
0 & {\rm otherwise}
\end{cases}
\label{findicator}
\end{equation}
In order to prove an inequality, the contraction $f$ must satisfy the boundary conditions 
\begin{equation}
f(x^{A_i}) = f^{A_i} \qquad {\rm and} \qquad f(x^{B_j}) = f^{B_j}
\label{bceq}
\end{equation}
for all $1 \leq i \leq m$ and all $1 \leq j \leq n$. These boundary conditions ensure that the RHS cuts implicitly defined by $f$ are in fact contiguous to (satisfy the homology condition for) the respective regions \cite{hec}. The boundary conditions are the only channel by which the contraction $f$ `knows about' the inequality it proves. 

\medskip
{\bf Conventions}~~
It is important to establish a convention for presenting the bits in $x \in \{0,1\}^l$ and $f(x) \in \{0,1\}^r$. We take `LHS terms' and `RHS terms' in definitions~(\ref{xindicator}) and (\ref{findicator}) to mean the terms that appear in:
\begin{equation}
\sum_{i=1}^m \sum_{j=1}^n S_{A_i^+ B_j^-}
\geq 
\sum_{i=1}^m \sum_{j=1}^n S_{A_i^- B_j^-} \,+ S_{A_1A_2\ldots A_m}
\tag{\ref{toricineqs}}
\end{equation}
Because we work with pure states on $m+n$ regions, each term can be traded for a complementary term without affecting the validity of the inequality. For example, inequality
\begin{equation*}
\sum_{i=1}^m \sum_{j=1}^n S_{A_i^+ B_j^-}
\geq 
\sum_{i=1}^m \sum_{j=1}^n S_{A_i^+ B_j^+} \,+ S_{B_1B_2\ldots B_n} 
\end{equation*}
is identical to (\ref{toricineqs}). Such switches affect the indicators defined in (\ref{xindicator}) and (\ref{findicator}), so that discussing contraction maps must always be done with respect to a particular presentation. 

Our convention differs from the convention employed elsewhere, in which one designated purifier region $O$ has preset indicators $x^O = \vec{0}$ and $f^O = \vec{0}$. That convention is motivated by a property of holographic inequalities called superbalance \cite{superbalance}. However, it is impractical for proving the toric inequalities. Instead, we stick to presentation~(\ref{toricineqs}). 

\subsubsection{Contraction map}
{\bf Definition}~~For each $x \in \{0,1\}^l$, define a graph $\Gamma(x)$ by drawing a horizontal/vertical pair of line segments on every square that carries a 0/1. This is illustrated in Figure~\ref{fig:gamma} in the main text. We explained there that $\Gamma(x)$ is a collection of nonintersecting loops, so it partitions the torus into connected components. 

Map $f$ is defined as follows. It assigns 1 to one special vertex in every connected component, which does not wrap a nontrivial cycle of the torus. We select that one special vertex to be the bottommost and, in case of ties, the rightmost vertex in the connected component, as displayed in Figure~\ref{fig:gamma}. The designations `bottommost' and `rightmost' are well-defined on components, which do not wrap nontrivial cycles of the torus. To all other vertices $f$ assigns 0. The value of $f$ in the last bit (on the distinguished RHS term $S_{A_1 A_2 \ldots A_m}$) is set so that:
\begin{equation}
|x|_1 \equiv |f(x)|_1 \quad \textrm{(mod 2)}
\label{toricparity}
\end{equation}

\medskip
{\bf Useful facts}~~
The loops, which comprise $\Gamma(x)$, are nonintersecting. The intersection number of two loops on a torus whose wrapping numbers around fundamental cycles are $(p_1, q_1)$ and $(p_2, q_2)$ is: 
\begin{equation}
\# = \left|\,\det \left( \begin{array}{cc} p_1 & p_2 \\ q_1 & q_2 \end{array} \right) \right|
\end{equation}
(Loops can have further intersections that are removable by deformation, but this number is a lower bound.) This number can only be zero if $(p_1, q_1) = (p_2, q_2)$ or if one of the $(p_i, q_i) = (0,0)$. We need not worry about the possibility that $(p_2, q_2)$ is a higher multiple of $(p_1, q_1)$, or vice versa, because such loops would self-intersect, which is also not allowed. Therefore, $p_1, q_1$ are relatively prime. 

We conclude that the wrapping number of every loop in $\Gamma(x)$ is either $(0,0)$ or $(p_1, q_1)$. Therefore, in talking about a single graph $\Gamma(x)$, we do not lose any information by representing loops simply as wrapped (1) or unwrapped (0). We adopt this convention in what follows.

\begin{figure}[t!]
    \centering
    \includegraphics[width=1\linewidth]{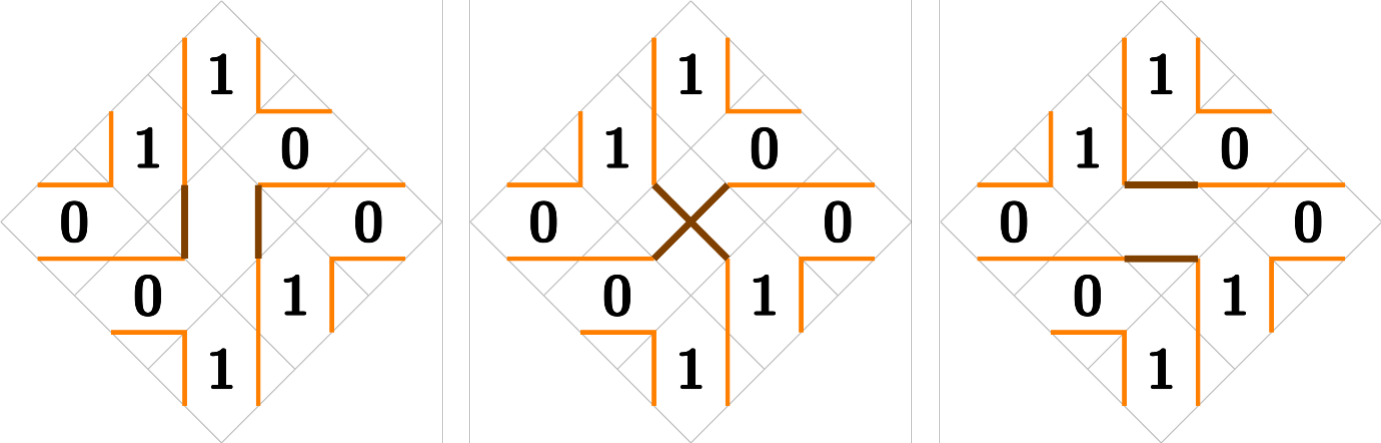}
    \caption{The case where a bit flip $x \to x'$ does not change the number of loops, illustrated on the $(m,n) = (3,3)$ torus. The graphs on the sides are $\Gamma(x)$ and $\Gamma(x')$ while the graph in the middle is $\tilde\Gamma$.}
    \label{fig:sameline}
\end{figure}

We want to study the effect of bit flips on $f(x)$ and, by extension, on $\Gamma(x)$. A bit flip (or the reverse process) breaks two neighboring loops in $\Gamma(x)$ and reconnects the resulting endpoints. If this process joins two distinct loops into one loop then its effect on the looping numbers is:
\begin{align}
0 + 0 & \leftrightarrow 0 \nonumber\\
0 + 1 & \leftrightarrow 1 \label{3effects} \\
1 + 1 & \leftrightarrow 0 \nonumber
\end{align}
It is also possible that the `two loops' which are being joined are, in fact, one and the same loop; see Figure~\ref{fig:sameline}. This scenario, which preserves the overall number of loops, is not covered in list~(\ref{3effects}) and requires a closer look. 

Consider a bit flip $x \to x'$ such that the total numbers of loops in $\Gamma(x)$ and in $\Gamma(x')$ are equal. This means that the bit flip transforms a single loop in $\Gamma(x)$ into a single loop in $\Gamma(x')$. We now argue that neither the initial loop in $\Gamma(x)$ nor the final loop in $\Gamma(x')$ can have the trivial wrapping $(0,0)$ on the torus. (Note that we are now contemporaneously discussing loops in two distinct graphs---$\Gamma(x)$ and $\Gamma(x')$---and therefore the bidivision into wrapped (1) and unwrapped (0) loops does not apply. That division referred to loops in a single graph $\Gamma(x)$.)

Let the wrapping numbers of the initial loop in $\Gamma(x)$ be $(p_x, q_x)$. Similarly, let the wrapping numbers of the final loop in $\Gamma(x')$ be $(p_{x'}, q_{x'})$. Now consider a third graph $\tilde\Gamma$, which differs from $\Gamma(x)$ and $\Gamma(x')$ only in the facet undergoing the flip. Instead of the connections featured there, it contains a cross that connects midpoints of opposite edges in the tiling; see Figure~\ref{fig:sameline}. Comparing $\tilde\Gamma$ to $\Gamma(x)$, we see that the initial loop with wrapping $(p_x, q_x)$ is replaced by two distinct loops in $\tilde\Gamma$, which intersect at one point. If we call the wrapping numbers of these two $\tilde\Gamma$-loops $(p_1, q_1)$ and $(p_2, q_2)$, we must have:
\begin{equation}
\det \left( \begin{array}{cc} p_1 & p_2 \\ q_1 & q_2 \end{array} \right) = \pm 1
\label{intersection1}
\end{equation}
Similarly, comparing $\tilde\Gamma$ to $\Gamma(x')$, we see that the graphs  differ only in the replacement of loops
\begin{equation}
(p_1, q_1),~(p_2, q_2) \to (p_{x'}, q_{x'})
\end{equation}

It is easy to relate the wrapping number of the initial loop $\Gamma(x)$ (respectively, the final loop in $\Gamma(x')$) to the two loops in $\tilde\Gamma$:
\begin{equation}
(p_x, q_x)~{\rm and}~(p_{x'}, q_{x'}) = (p_1, q_1) \pm (p_2, q_2) 
\label{3rdcross}
\end{equation}
This follows by a simple deformation argument. But quantity~(\ref{3rdcross}) cannot be $(0,0)$ if (\ref{intersection1}) holds. This shows that every bit flip, which preserves the overall number of loops, destroys and creates a nontrivially wrapped loop.  

\begin{figure}[t]
    \centering
    \includegraphics[width=0.86\linewidth]{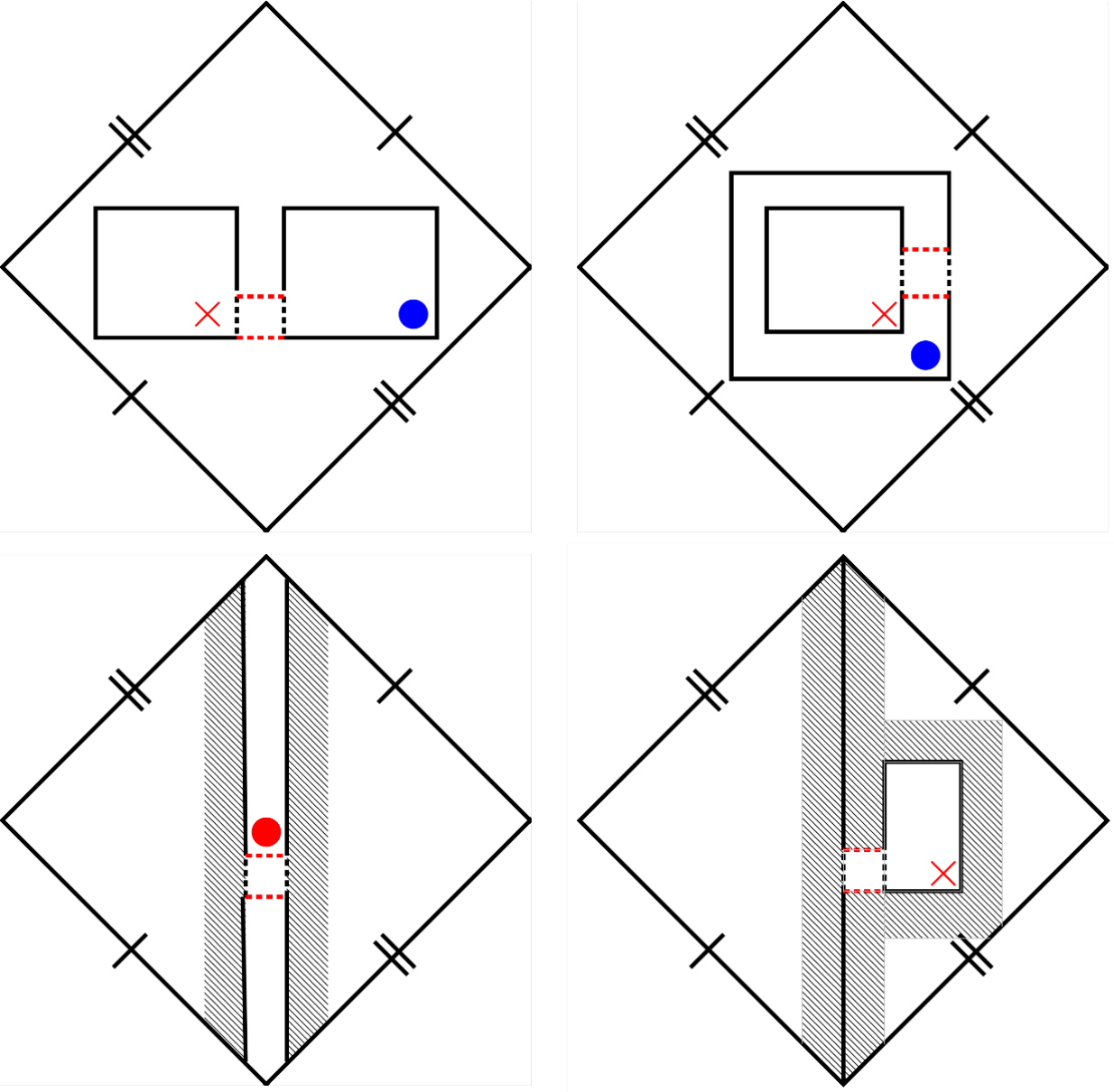}
    \caption{
 A bit flip $x \to x'$, which joins two loops in $\Gamma(x)$ in the proof of toric inequalities. Black dotted lines are destroyed and red dotted lines are created.  There are four distinct cases, which come from list~(\ref{3effects}), with $0 + 0 \leftrightarrow 0$ further subdivided into two subcases. The two subcases differ in whether the two merged loops form the boundary of an annulus or two disks. (They cannot do both because then they would live on a sphere instead of a torus.) \\
Vertices, on which $f$ flips are marked in red (X for $1 \to 0$ flips; circle for $0 \to 1$ flip). Vertices, on which $f$ remains at 1 are marked with blue circles. Shaded regions are locked at 0 because they wrap a nontrivial cycle of the torus.
    }
    \label{fig:joiningloops}
\end{figure}

\medskip
{\bf Proof that $f$ is a contraction}~~Based on lemma~(\ref{simplification}), we wish to study the effect of flipping a single bit in $x$ on $f(x)$. Because $f(x)$ is defined in terms of connected components on the torus, we wish to study how a bit flip impacts the said components. We can do so by analyzing the effect of a bit flip on the wrapping numbers of loops in $\Gamma(x)$. This is because the loops of $\Gamma(x)$ form boundaries of the connected components. Our analysis uses this obvious fact:
\begin{itemize}
\item If the boundary of a connected component wraps a nontrivial cycle of the torus then the component itself wraps the same cycle. 
\item Contrapositive: If a component does not wrap a nontrivial cycle then its boundary consists entirely of unwrapped loops. These components are important because they---and only they---contain vertices (one per component) to which map $f$ assigns 1. 
\end{itemize}
We now inspect what happens to the connected components on the torus in each bit flip scenario listed above.

The three cases in which a bit flip joins two loops into one (or the reverse) are listed in (\ref{3effects}). We display them in Figure~\ref{fig:joiningloops}. In each of them, the value of $f(x)$ changes at exactly one vertex in the tiling. In agreement with condition~(\ref{toricparity}) the $r^{\rm th}$ bit in $f(x)$ does not change. Overall, we find that every bit flip $x \to x'$, which changes the number of loops in $\Gamma(x)$, results in
\begin{equation*}
|f(x) - f(x')|_1 = 1
\end{equation*}
with the $r^{\rm th}$ bit preserved. 

Now consider the case in which a bit flip in $x$ preserves the number of loops in $\Gamma(x)$. Then the region contiguous to the initial loop in $\Gamma(x)$ is the same as the region contiguous to the final loop in $\Gamma(x')$; see Figure~\ref{fig:sameline}. (We leave it as an exercise for the reader to show that both sides of the created/destroyed loop actually live in one connected component on the torus, which is why we say `region' instead of `regions.') Moreover, the region is nontrivially wrapped around the torus so all the tiling vertices therein are set to 0 both under $f(x)$ and under $f(x')$. Since $f(x)$ does not change on any tiling vertices, condition~(\ref{toricparity}) implies that the $r^{\rm th}$ bit in $f(x)$ flips. Overall, we find that every bit flip $x \to x'$, which does not change the number of loops in $\Gamma(x)$, results in
\begin{equation*}
|f(x) - f(x')|_1 = 1
\end{equation*}
with the $r^{\rm th}$ bit flipped. 

In summary, we have verified
\begin{equation}
|x-x'|_1 = 1 \quad \Rightarrow \quad |f(x) - f(x')|_1 = 1
\label{toricfinalproof}
\end{equation} 
so $f$ is a contraction.

\subsubsection{Boundary conditions}
\begin{figure}[t]
		\centering
		\includegraphics[width=1\linewidth]{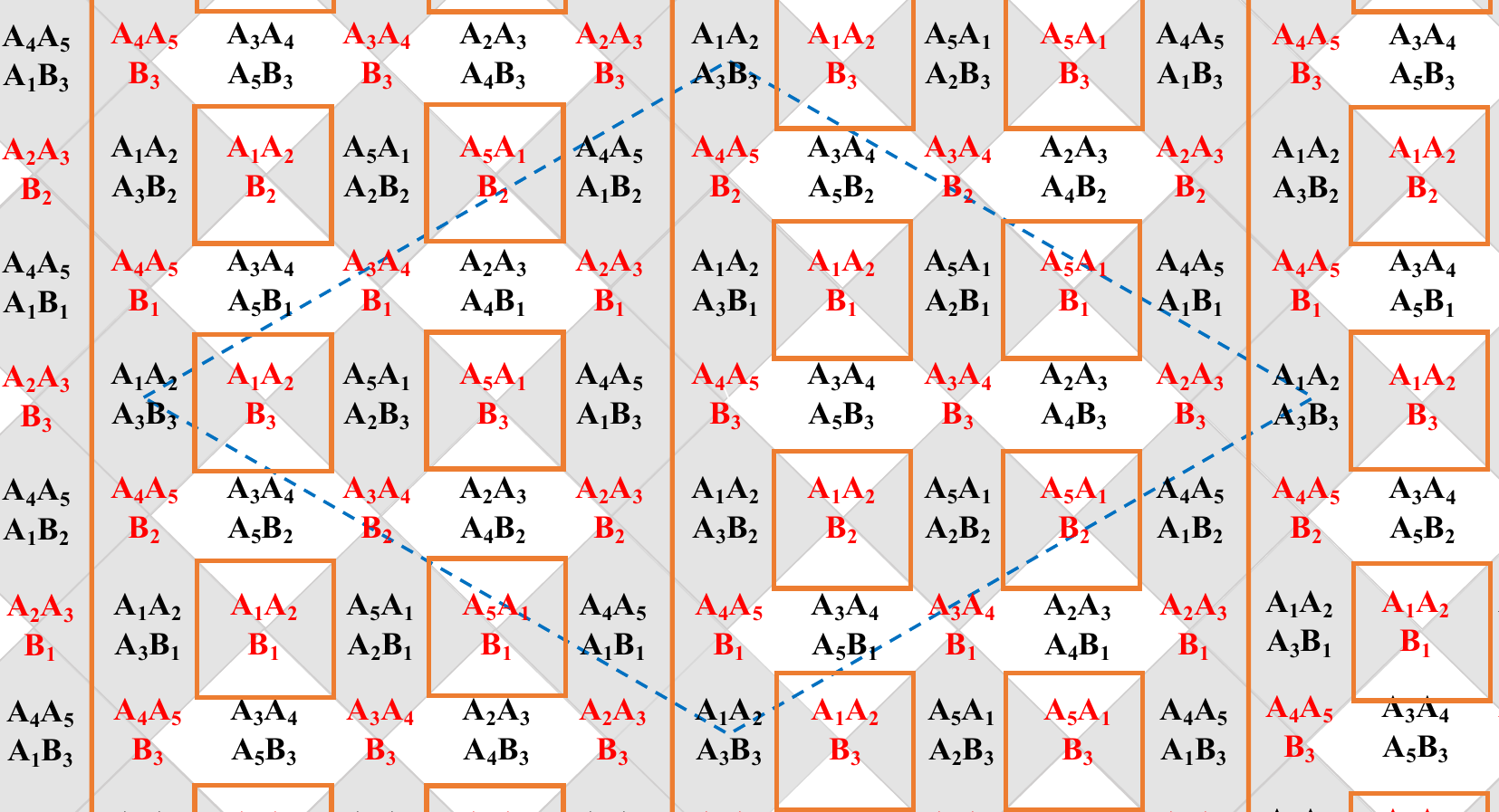} \\
		\vspace*{2mm}
		\includegraphics[width=1\linewidth]{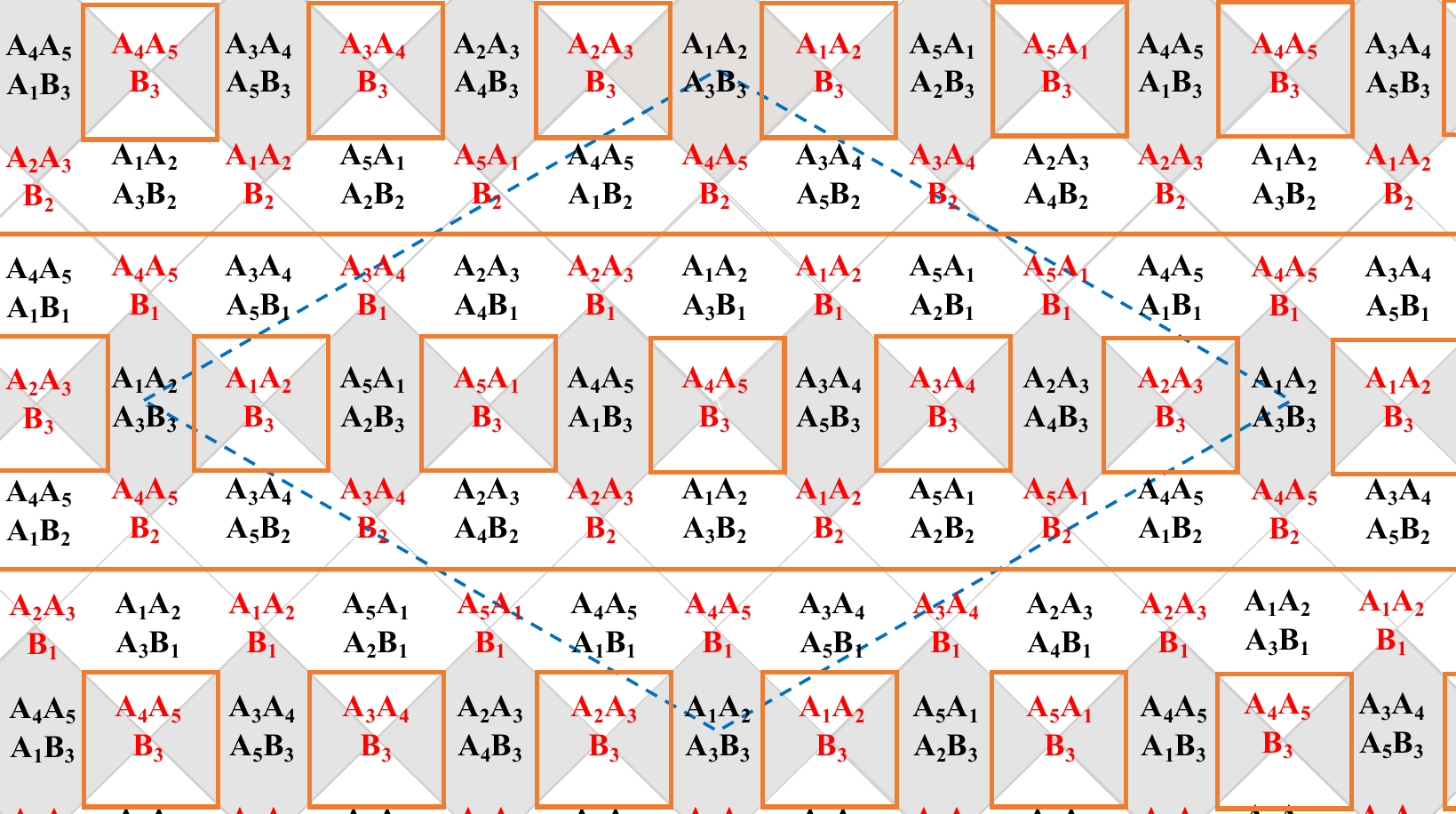}
	\caption{Boundary condition for $A_1$ (upper panel) and for $B_3$ (lower panel) in the $(m,n) = (5,3)$ toric inequality. Squares, which contain $A_1$ (respectively $B_3$), that is squares carrying 1's in the indicators $x^{A_1}$ and $x^{B_3}$, are shaded. Graphs $\Gamma(x^{A_1})$ and $\Gamma(x^{B_3})$, shown in orange, surround with loops precisely those vertices of the tiling, which contain $A_1$ (respectively $B_3$). These vertices---and only they---are mapped to 1 under $f$, in agreement with boundary condition~(\ref{bceq}).}
	\label{fig:bcs}
\end{figure}

Figure~\ref{fig:bcs} shows $\Gamma(x^{A_1})$ and  $\Gamma(x^{B_3})$. On the tiling, the resulting $f(x^{A_1})$ and $f(x^{B_3})$ match $f^{A_1}$ and $f^{B_3}$. Since the construction is periodic in both directions on the torus, this conclusion extends to boundary conditions for all other regions. 

Regarding the $r^{\rm th}$ (non-tiling) bit of $f(x)$, we have:
\begin{align}
|x^{A_i}|_1 = n(m+1)/2 & = |f^{A_i}|_1 \nonumber \\
|x^{B_j}|_1 = m(n-1)/2 & = |f^{B_j}|_1
\end{align}
So the boundary conditions satisfy $|x|_1 \equiv |f(x)|_1$ (mod 2), which is the defining equation for the $r^{\rm th}$ bit of $f(x)$. 

This completes the proof of inequalities~(\ref{toricineqs}). 

\subsubsection{Remarks}
{\bf What makes the inequalities true?}~~
Inequalities~(\ref{toricineqs}) are only meaningful because the contraction $f$ can be extended from the faces and vertices of the tiling to an additional, $r^{\rm th}$ bit in $f(x)$. This bit is responsible for the special RHS term $S_{A_1 A_2 \ldots A_m}$. Without it, the inequalities would simply be linear combinations of strong subadditivity; see ineq.~(\ref{ineqcmi}) in the main text.

Therefore, it is interesting to inspect the meaning of the last bit in $f(x)$. The analysis that led to assertion~(\ref{toricfinalproof}) showed that the final bit in $f(x)$ flips if and only if the bit flip $x \to x'$ preserves the number of loops in $\Gamma(x)$. After inspecting the special case $x = \vec{0}$, we may write a closed formula for this bit:
\begin{equation}
(f(x))_r = |x|_1 - \# \{\textrm{loops in}~\Gamma(x)\} + 1\quad \textrm{(mod 2)}~~
\end{equation}
Evidently, this $\mathbb{Z}_2$-valued index is the topological property, which $f(x)$ keeps track of and which supports the term $S_{A_1 A_2 \ldots A_m}$ in the toric inequalities. 

\medskip
{\bf Boundary conditions, differential entropy}~~The topologically nontrivial loop, which appears in $\Gamma(x^{B_3})$ in Figure~\ref{fig:bcs}, crosses precisely those edges of the tiling, which had ${\rm CMI}_{\rm edge} \neq 0$ in equation~(\ref{continuumintegrand}) in the main text. That equation referred to the situation where $B_n$ is so large that it alone decides the phases of all terms $S_{A_i^\pm B_j^-}$. In this circumstance, the inequality reduced to the form:
\begin{equation}
\oint dv \frac{\partial S(u,v)}{\partial v} \Big|_{u = u(v)} 
\geq S_{A_1 A_2 \ldots A_m}
\label{sdiffagain}
\end{equation}
Viewed on the torus, the path of integration in (\ref{sdiffagain}) is precisely the wrapping loop in $\Gamma(x^{B_n})$.

\subsection{Proof of projective plane inequalities}
\subsubsection{Generalities}
{\bf Convention}~~Before proving the inequalities, we must set the convention for how each term is written down (whether or not it is traded for its complement). 

We start with LHS terms. In (\ref{app:rp2ineqs}), they are double-counted and then corrected by a factor of $1/2$; we will not do that in this proof. It is possible to write every LHS term singly, but at the cost of breaking the symmetry between:
\begin{align}
S_{A_i^{(j)} B_{i+j}^{(m-j)}} & = S_{A_{i+j}^{(m-j)} B_{i}^{(j)}} \nonumber \\
S_{A_i^{(j)} B_{i+j-1}^{(m-j)}} & = S_{A_{i+j}^{(m-j)} B_{i-1}^{(j)}} \label{appchoices}
\end{align}
We did so, for example, in going from (\ref{app:rp2ineqs}) to (\ref{rp2asymm}), but one could make other choices and rewrite the $\mathbb{RP}^2$ inequalities in ways other than (\ref{rp2asymm}).

\begin{figure}[t]
\centering
\includegraphics[width=0.99\linewidth]{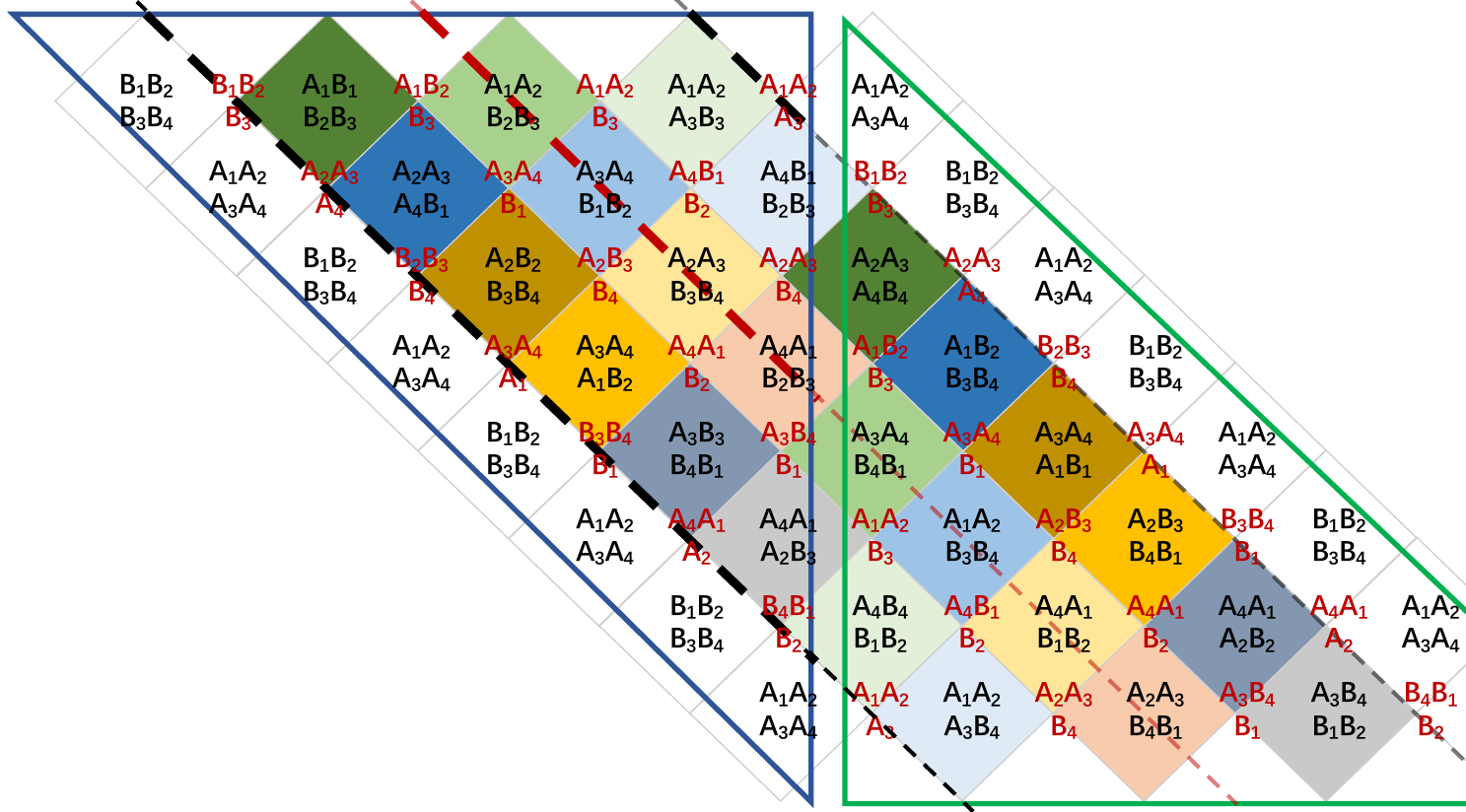}
\caption{Our proof is easiest to understand when we represent each LHS term the way it appears inside one fundamental domain shown here. The dashed red line (equator) and the black dashed line (polar circle) are explained in Appendix~\ref{app:rp2proof}.}
 \label{fig:rp2proof}
 \end{figure}

It will turn out that the choices encountered in (\ref{appchoices}) do not matter for our definition of the contraction map. Nevertheless, it is visually helpful to have in mind a definite choice. To specify it, we refer to Figure~\ref{fig:rp2proof}. It is the same as the $\mathbb{RP}^2$ panel in Figure~\ref{fig:examples} in the main text, except for the highlighted choice of fundamental domain under the identification 
\begin{equation}
\textrm{M{\"o}bius strip} 
= \frac{\textrm{infinite strip}}{\textrm{translation by $m$ squares} \times \textrm{a flip}}.
\label{quotient}
\end{equation}
A key feature of Figure~\ref{fig:rp2proof} is that distinct choices described in (\ref{appchoices}) are related precisely by the symmetry `translation by $m$ squares followed by a flip,' which is quotiented out in equation~(\ref{quotient}). Therefore, making a choice in (\ref{appchoices}) can be thought of as choosing one fundamental domain of identification~(\ref{quotient}). For our purposes, it is most transparent (though not logically necessary) to work with the choice defined by the fundamental domain in Figure~\ref{fig:rp2proof}. 

We take all RHS terms to be written in their $(m-1)$-partite form $S_{A_i^{(j-1)} B_{i+j-1}^{(m-j)}}$ rather than the complementary $(m+1)$-partite form.

In summary, for the purposes of defining and studying a contraction, we write inequalities~(\ref{app:rp2ineqs}) in the form:
\begin{align}
(m-1)\, S_{A_1A_2 \ldots A_m} & \nonumber \\
+ \sum_{j=1}^{\myfloor{m/2}} \!\! \sum_{i=1}^{~i_{\rm max}(j)}
& \left(S_{A_i^{(j)} B_{i+j-1}^{(m-j)}}~~{\rm or}~~S_{A_{i+j}^{(m-j)} B_{i-1}^{(j)}}\right) \nonumber \\
+ \sum_{j=1}^{\myfloor{m/2}} \!\! \sum_{i=1}^{~i_{\rm max}(j)}
& \left(S_{A_i^{(j)} B_{i+j}^{(m-j)}}~~{\rm or}~~S_{A_{i+j}^{(m-j)} B_{i}^{(j)}}\right) \nonumber \\
\geq 
\sum_{j,i=1}^m & S_{A_i^{(j-1)} B_{i+j-1}^{(m-j)}}
\label{rp2asymm2}
\end{align}
In each expression $(\,\ldots~{\rm or}~\ldots)$ we take the term, which lives in some chosen fundamental domain of quotient~(\ref{quotient}). The choice of fundamental domain does not affect the proof, but it affects the readability of the proof. The proof is easiest to absorb when it is visualized on the fundamental domain shown in Figure~\ref{fig:rp2proof}. 

\medskip
{\bf Adjusted lemma}~~Our proof of the toric inequalities used lemma~(\ref{simplification}). But in the presence of unequal coefficients, the distance $|x - x'|_1$ weighs bitwise differences between $x$ and $x'$ by their coefficients in the inequality. As stated, lemma~(\ref{simplification}) does not allow the possibility that flipping a single physical bit on the LHS might result in $|x - x'|_1$ greater than one. We need to extend the lemma in the obvious way. 

The logic of the lemma is that for every pair $x(0), x(k) \in \{0,1\}^l$ such that $|x(0) - x(k)|_1 = k$ we can consider a trajectory of single bit flips $x(0) \to x(1) \to x(2) \ldots \to x(k)$. If a single bit flip is contracting then so is their sequence. We now rewrite the lemma, but allowing for the possibility that $x(s)$ and $x(s+1)$ may differ in only one physical bit yet their distance is greater than one. (Alternatively, one could treat identical LHS terms as distinct but unphysical bits; here we stick to physically meaningful bits instead.) In inequalities~(\ref{app:rp2ineqs}) this happens only in the term $S_{A_1 A_2 \ldots A_m}$ where the distance is $m-1$. In summary, the modified form of the lemma is:
\begin{align}
{\rm if}~|f(x) - f(x')|_1 & \leq |x-x'|_1~~\forall~|x-x'|_1 = 1 \nonumber \\
{\rm and}~|f(x) - f(x')|_1 & \leq m-1 \quad \forall~x,x'~\textrm{that differ} \nonumber \\
& \phantom{\leq m-1 \quad \forall}~~\textrm{in $S_{A_1 A_2 \ldots A_m}$ only} \nonumber \\
{\rm then}~|f(x) - f(x')|_1 & \leq |x-x'|_1 \quad \textrm{for all}~~x, x'
\label{rp2simplification}
\end{align}

\subsubsection{Contraction map}
\label{app:rp2deff}
{\bf Preliminaries}~~We define a contraction $x \to f(x)$ in a manner, which is similar to the proof of the toric inequalities. Before that, a few remarks specific to the settings of the projective plane are in order.

With reference to Figure~\ref{fig:rp2proof}, we represent every bit in $x \in \{0,1\}^l$ as a pair of horizontal (0) or vertical (1) line segments, which connect midpoints of sides of the relevant square; see Figure~\ref{fig:gamma} in the main text for illustration. This produces a graph, which is almost the graph we want. We now modify the graph slightly; the structure resulting from the modification will be called $\Gamma(x)$.

Let us concentrate on the boundary diagonals (leftmost and rightmost) in Figure~\ref{fig:rp2proof}. (Here `boundary' means boundary of the infinite strip, which is quotiented into a M{\"o}bius strip.) Each fundamental domain in Figure~\ref{fig:rp2proof} contains $2m$ squares on one boundary diagonal and zero squares on the other boundary diagonal. For concreteness, we consider the fundamental domain highlighted in blue, which encompasses squares on the leftmost diagonal. Of the $2m$ boundary squares there, $m$ are labeled $A_1 A_2 \ldots A_m$ and the other $m$ are labeled with the complement $B_1 B_2 \ldots B_m$. (It is possible to rationalize why we have $2m$ such squares despite the inequality having only $m-1$ copies of $S_{A_1 A_2 \ldots A_m} = S_{B_1 B_2 \ldots B_m}$. This interpretive exercise does not affect the validity of the proof, so we leave it as homework for the interested readers.) 

Observe that the labels on the boundary diagonal alternate between $A_1 A_2 \ldots A_m$ and $B_1 B_2 \ldots B_m$. Therefore, for every $x \in \{0,1\}^l$, the squares on the boundary diagonal are filled with an alternating pattern of 0s and 1s. On such a configuration, our rules produce an alternating pattern of ticks, which look like $\rightangle \rotatebox[origin=c]{180}{$\rightangle$}\, \rightangle \rotatebox[origin=c]{180}{$\rightangle$}$. Moreover, on the leftmost diagonal, the ticks of type $\rotatebox[origin=c]{180}{$\rightangle$}$ are isolated; they rest on the boundary of the strip and do not extend into loops. (If we worked on the other fundamental domain in Figure~\ref{fig:rp2proof}, the same characterization would apply to ticks $\rightangle$ on the rightmost diagonal.) We remove such isolated ticks from the graph. After the removal, the resulting graph is called $\Gamma(x)$. 

\begin{figure}[t]
\centering
\includegraphics[width=0.99\linewidth]{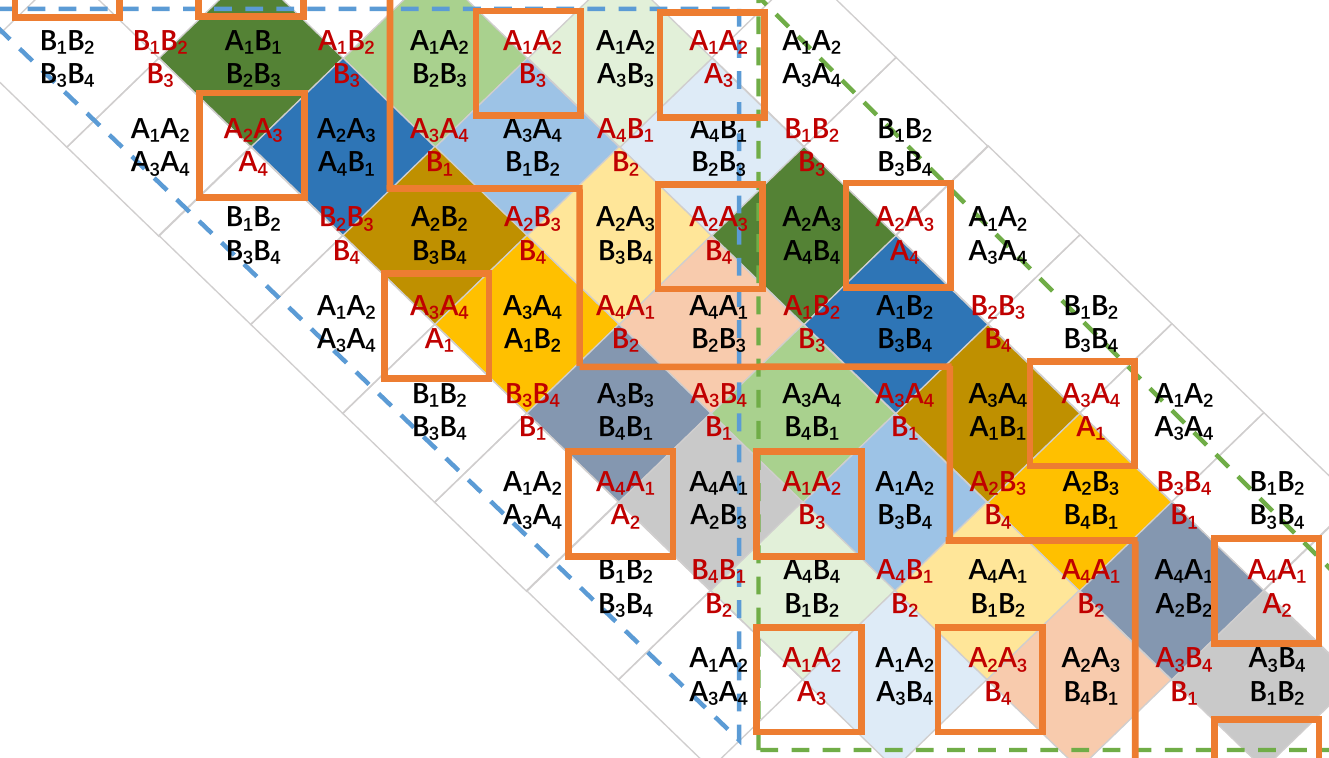}
\caption{An example graph $\Gamma(x)$. Note the noncontractible loop, which runs through the middle of the strip.}
 \label{fig:rp2example}
 \end{figure}

A key property of $\Gamma(x)$ is that it consists of nonintersecting loops. As in the toric case, the loops of $\Gamma(x)$ partition the underlying manifold $\mathbb{RP}^2$ into connected components. An example graph $\Gamma(x)$ is shown in Figure~\ref{fig:rp2example}.

To define $f(x)$, we must place 0's and 1's on tiling vertices (corners of squares) in Figure~\ref{fig:rp2example}. We exclude the corners on the boundary of the strip because they do not correspond to RHS terms of the inequality. (In the preceding paragraph, we removed the ticks $\rotatebox[origin=c]{180}{$\rightangle$}$ from graph $\Gamma(x)$ partly because they do not make up loops, but also because in our construction they could only be relevant to the meaningless boundary corners.)

In the proof of the toric inequalities, $f(x)$ assigned 1 to the bottommost and rightmost vertex in each connected component of the torus, which does not wrap a non-contractible cycle of the torus. We would like to do something similar here, but the fact that $\mathbb{RP}^2$ is not orientable renders the description `bottommost and rightmost' meaningless. To solve this problem, we introduce a priority ranking on the vertices. That is, we list all vertices (RHS terms) in some arbitrary sequence and label them $v_1, v_2, \ldots$ up to $v_{m^2}$. (The largest index is $m^2$ because inequality~(\ref{rp2asymm2}) has $m^2$ terms on the RHS.) Instead of `bottommost and rightmost vertices,' we will be designating `top priority vertices,' where top priority means lowest index in a given collection $\{v_i\}$ of RHS terms. 

\medskip
{\bf Definition of contraction}~~We define map $x \to f(x)$ as follows:

For every $x \in \{0,1\}^l$, construct graph $\Gamma(x)$ as above and use it to divide $\mathbb{RP}^2$ into connected components. Map $f(x)$ assigns 1 to the top priority vertex in each connected component, which does not wrap a nontrivial cycle of $\mathbb{RP}^2$. All other vertices are assigned 0.

\subsubsection{Boundary conditions}
\begin{figure}[t]
\centering
\includegraphics[width=0.99\linewidth]{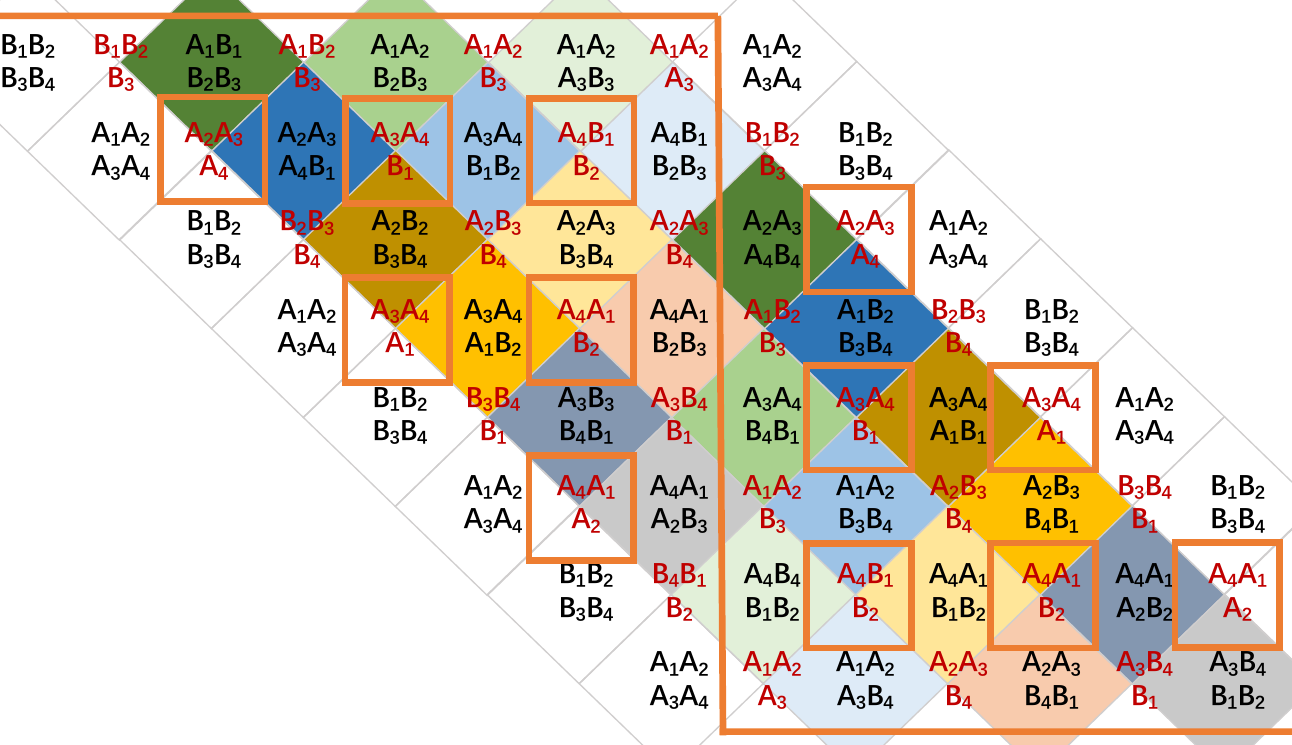}
\caption{Graph $\Gamma(x^{A_4})$, which passes vertically (respectively horizontally) through squares whose labels contain (exclude) $A_4$. Vertices that are surrounded by the small loops (and only they) are assigned 1 in $f(x^{A_4})$. They are precisely the vertices, which contain $A_4$ when written in their $(m-1)$-partite form. This means $f(x^{A_4}) = f^{A_4}$ and the boundary condition for $A_4$ is satisfied. Under the symmetry $D_{2m}$ that preserves the inequality, the orbit of $A_4$ covers all regions. The construction of $f$ respects this symmetry, so all boundary conditions are satisfied.}
 \label{fig:rp2bc}
 \end{figure}

We check the boundary conditions first because they will be useful in establishing that $f$ is a contraction.  We verify the boundary condition for region $A_4$ explicitly in Figure~\ref{fig:rp2bc}. This automatically verifies the boundary conditions for all regions because our construction respects the symmetry $D_{2m}$ whose action sends $A_4$ to all other regions.

\subsubsection{Useful facts for proving that $f$ is a contraction}
\label{app:rp2proof}

\medskip
{\bf Facts about $\mathbb{RP}^2$}~~The real projective plane has one noncontractible cycle and the fundamental group of $\mathbb{RP}^2$ is $\mathbb{Z}_2$. In particular, any loop in $\mathbb{RP}^2$ either does (1) or does not (0) wrap the noncontractible cycle. When two such loops merge, they obey the rules in (\ref{3effects}), which is simply the group law of $\pi_1(\mathbb{RP}^2) = \mathbb{Z}_2$.

We will need to pay attention to whether or not the loops that make up $\Gamma(x)$ are contractible on $\mathbb{RP}^2$. In doing so, we will use intersection numbers of loops on $\mathbb{RP}^2$:
\begin{equation}
\#(0,0) = \#(0,1) = 0 \qquad {\rm and} \qquad \#(1,1) = 1
\label{intersects}
\end{equation}
The vanishing ones are obvious because the trivial loop can be contracted to a point. The nonvanishing intersection of the wrapped loop with itself can be easily understood on the two-sphere, which is a double cover of $\mathbb{RP}^2$. Any two great circles intersect at two antipodal points on the sphere. Under the quotient $S^2 \to \mathbb{RP}^2$, the great circles descend to noncontractible loops on $\mathbb{RP}^2$ and their two intersection points become one point.

Specifically, we will utilize a test loop on $\mathbb{RP}^2$, which is marked in dashed red in Figure~\ref{fig:rp2proof}. The test loop is not part of any graph $\Gamma(x)$; it is only used to test the wrapping of constituent loops in $\Gamma(x)$. The property that makes this loop useful for testing other loops is that it is noncontractible. (Any other noncontractible loop would serve as a test loop just as well.) The test loop we chose is nice because it runs in the middle of the strip, so we can describe it in a uniform way for all $m$. Note, however, that for even $m$ our test loop runs through the squares on the middle diagonal while for odd $m$ there is no middle diagonal and the test loop runs instead on edges of the tiling. The symmetric location of the test loop motivates calling it the equator. 

The intersection numbers~(\ref{intersects}) imply that we can easily diagnose the contractibility of any loop. If a loop intersects the equator in an odd number of points---one from $\#(1,1) = 1$ plus possibly some removable intersections, which come in pairs---then it is noncontractible. If a loop intersects the equator in an even number of points---all in removable pairs---then it is contractible. 

Finally, we note for future use that $\mathbb{RP}^2$ can be represented as a sphere with a single cross-cap inserted. 

\medskip
{\bf Every $\Gamma(x)$ has a noncontractible loop}~~If it had more than one then they would intersect, so the nontrivial claim is that it contains at least one. To check the claim in an example, see Figure~\ref{fig:rp2example}.

The first step in verifying the claim is to inspect $\Gamma(x^{A_4})$ in Figure~\ref{fig:rp2bc}. It contains a noncontractible loop, which zigzags between the leftmost and the rightmost diagonals and crosses the equator. Every fundamental domain contains one such crossing (an odd number) and therefore the loop is noncontractible. 

We now analyze what happens when a bit of $x$ is flipped. First, consider $x \to x'$ related by an interior bit flip, that is one not on the boundary diagonals. If this flip combines two loops into one (or vice versa) then its effect on the loops of $\Gamma(x)$ is captured by the $\pi_1(\mathbb{RP}^2) = \mathbb{Z}_2$ group law (\ref{3effects}). That group law can only create or destroy noncontractible loops in pairs. Therefore, if $\Gamma(x)$ contains an odd number of noncontractible loops (that is, one) then so does $\Gamma(x')$.

Second, we consider an interior bit flip $x \to x'$, which preserves the number of loops. In fact, this scenario cannot happen at all, but it is easier to establish this as a corollary after we know that $\Gamma(x)$ contains a noncontractible loop. We now argue that if an interior bit flip could turn one loop in $\Gamma(x)$ into a single distinct loop in $\Gamma(x')$ then they would both have to be contractible. Therefore, this type of bit flip---even if it could happen---would not create or remove noncontractible loops.

To see that, we adapt the argument around equations~(\ref{intersection1}-\ref{3rdcross}) from the toric proof. Suppose a bit flip $x \to x'$ turns a single loop in $\Gamma(x)$ into a single loop in $\Gamma(x')$. Construct a new graph $\tilde\Gamma$, in which the square undergoing the flip contains a cross, like in the middle picture in Figure~\ref{fig:sameline}. By construction, $\tilde\Gamma$ contains two loops $l_1$ and $l_2$, which intersect at one point, so they are both noncontractible. The initial loop in $\Gamma(x)$ and the final loop in $\Gamma(x')$ are obtained by combining $l_1$ with $\pm l_2$, so their wrapping numbers are $1 \pm 1 = 0$. This establishes that a bit flip, which preserves the overall number of loops, acts on their wrapping numbers as $0 \to 0$. If such flips occurred, they would not help in removing a noncontractible loop from $\Gamma(x)$. 

Finally, we consider a flip $x \to x'$ on the boundary diagonals in Figure~\ref{fig:rp2proof}. This operation happens away from the equator so it does not change the total number of intersections between $\Gamma(x)$ and the equator, which we know is odd. But if $\Gamma(x')$ has an odd total number of intersections with the equator, it must contain a noncontractible loop. Indeed, if $\Gamma(x')$ was a union of only contractible loops, that intersection number would be even.

We analyzed all possible bit flips $x \to x'$: in the interior region (flips that do / do not preserve the total count of loops) and on boundary diagonals. None of them can remove a noncontractible loop. Every $\Gamma(x)$ can be reached by some sequence of bit flips from $\Gamma(x^{A_4})$, which contains a noncontractible loop. Therefore so does every $\Gamma(x)$. 

We note a {\bf corollary}: loop number-preserving bit flips $x \to x'$ do not occur. We saw that if they did, they could only take the form $0 \to 0$ (meaning they involve contractible loops). We also saw that by putting a cross on the square undergoing the flip---a local operation---we could split the loop in question into two noncontractible loops $l_1$ and $l_2$ in a new graph $\tilde\Gamma$. Now use the fact that $\Gamma(x)$ already contains a noncontractible loop. That noncontractible loop necessarily intersects $l_1$ and $l_2$. But those intersections must also be there in $\Gamma(x)$ because away from the square undergoing the flip $l_1$ and $l_2$ are identical to the initial loop in $\Gamma(x)$. This contradicts the fact that $\Gamma(x)$ consists of mutually nonintersecting loops.

\medskip
{\bf Parity on the polar circle}~~We call `polar circle' the tiling line, which separates the boundary diagonal from the interior squares. It is marked in dashed black in Figure~\ref{fig:rp2proof}. The figure displays one dashed line by the leftmost diagonal and another by the rightmost diagonal, but the two are identified with one another. In the same nomenclature, we could call the boundary diagonals the polar zone or North/South pole. Note that the polar circle is contractible in $\mathbb{RP}^2$ because we could slide and contract it to a point on the North/South pole.

The polar circle comprises $2m$ edges, which can be sorted into `odd edges' and `even edges.' We call `odd' the edges on the polar circle that are incident to squares labeled $B_1 B_2 \ldots B_m$. The other edges, incident to squares with $A_1 A_2 \ldots A_m$, will be called `even.' Distinguishing even and odd edges on the polar circle is useful because of the following {\bf lemma}: Consecutive intersections of a loop $l \subset \Gamma(x)$ with the polar circle alternate between even and odd edges. 

\begin{figure}[t]
\centering
\includegraphics[width=0.99\linewidth]{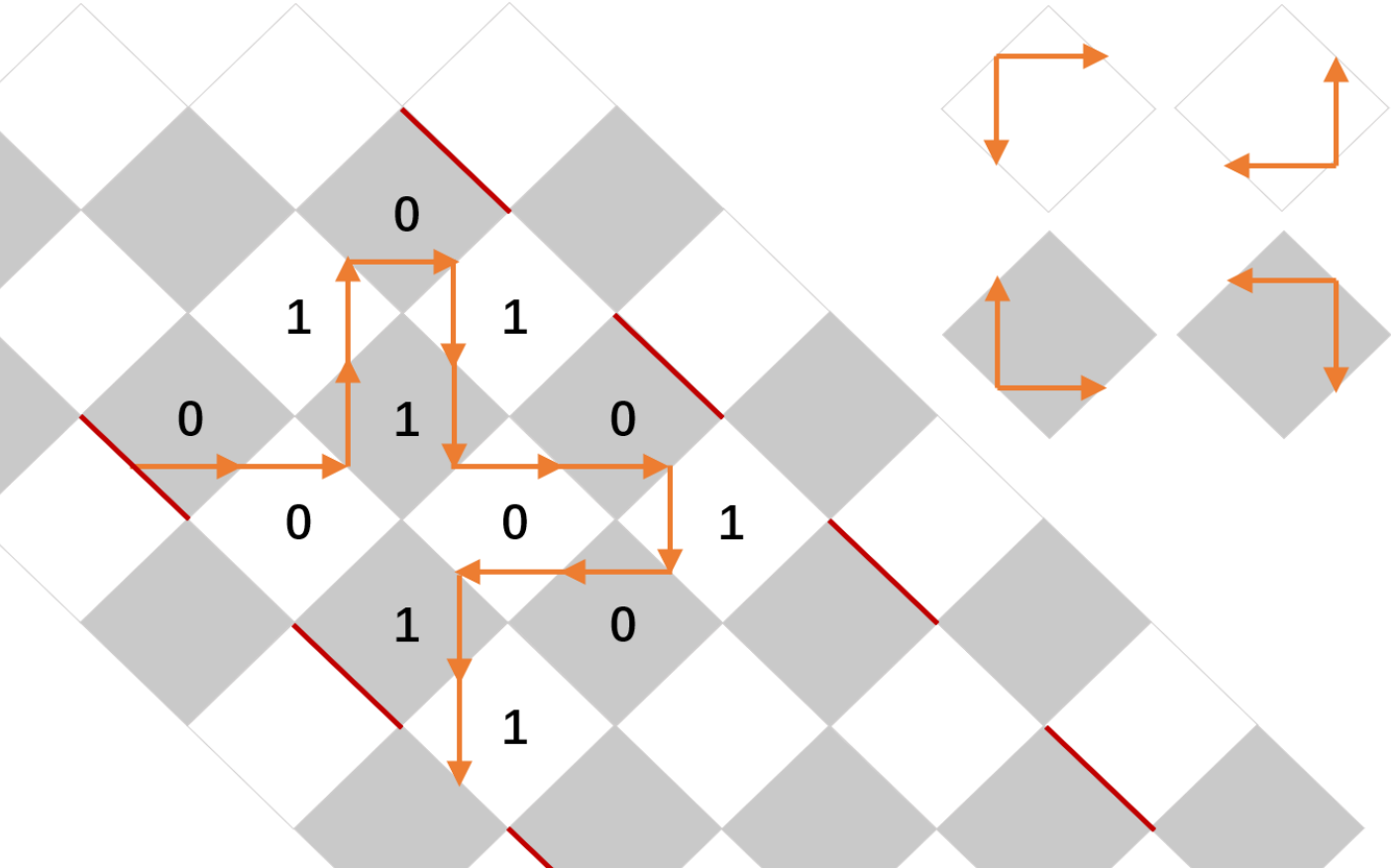}
\caption{The intersections of a loop with the polar circle alternate between even and odd edges. Once we pick a directionality on the loop, we see that the loop jumps from right-moving edges to left-moving edges only on light squares. Odd edges on the polar circle are highlighted.}
 \label{fig:parity}
 \end{figure}

To establish the lemma, color the squares in Figure~\ref{fig:rp2proof} like a chessboard; see Figure~\ref{fig:parity}. For definiteness, let the squares labeled $A_1 A_2 \ldots A_m$ be the dark squares and those labeled $B_1 B_2 \ldots B_m$ be the light squares. This way, odd polar circle edges have a light square to their left bottom and a dark square to their right top; for even edges it is the reverse. 

Consider the sequence of tiling edges visited by some loop $l \subset \Gamma(x)$. It is clear that they alternate between `left-moving' and `right-moving' edges in the obvious nomenclature. We initialize the sequence of edges visited by loop $l$ at an intersection with the polar circle. (If the loop does not intersect the polar circle then the claim is trivially true.) For definiteness, assume that initial intersection occurs on an odd edge. From there, let us follow the loop into the interior region---not out to the boundary diagonal. Because the initial edge is odd, the loop first traverses a dark square and jumps from a left-moving edge to a right-moving edge. In the next step, the loop traverses a light square and jumps from a right-moving edge to a left-moving edge. The pattern then repeats. Consider the next intersection of loop $l$ with the polar circle. (There must be an even number of intersections because the polar circle is contractible.) It will be reached by a jump from a right-moving edge to a left-moving edge because the polar circle is made up of left-moving edges. Such jumps traverse light squares, so the next intersection falls on an even edge as claimed.

\medskip
{\bf Loops of $\Gamma(x)$ and permutations}~~Take graph $\Gamma(x)$ and remove the boundary diagonal (the `polar zone'). The lemma we just established implies that in the resulting truncated graph odd edges are attached to even edges by line segments. 

Let us label the edges of the polar circle in cyclic order $(1_o, 1_e, 2_o, 2_e, \ldots m_o, m_e)$. Suppose edge $p_o$ is attached to edge $q_e$ in $\Gamma(x)\setminus \textrm{(polar zone)}$. Then we can think of $\Gamma(x)$ as defining a permutation $\pi(x) \in S_m$, which sends $p \to q$. 

In fact, $\pi(x)$ is a noncrossing permutation \cite{stanley}. To see this, we again start with the boundary conditions in Figure~\ref{fig:rp2bc} and observe that permutation $\pi(x^{A_4})$ is manifestly noncrossing. The claim then follows because bit flips $x \to x'$ reroute but never cross the line segments in $\Gamma(x)\setminus \textrm{(polar zone)}$, which define $\pi(x)$. 

We would like to relate the loops of $\Gamma(x)$ to $\pi(x)$. Here there are two possibilities, depending on the $S_{A_1 A_2 \ldots A_m}$ bit in $x$. As we discussed in Preliminaries in Appendix~\ref{app:rp2deff}, that bit contributes to $\Gamma(x)$ a series of ticks $\rightangle$. The ticks either encircle the vertices, which have an odd edge to the left and an even edge to the right of them, or vice versa. We interpret the ticks as setting up another permutation, which either sends $q_e \to q_o$ or $q_e \to (q+1)_o$. As permutations, these two options define the identity $e$ and the cyclic permutation $\xi = (1\, 2\, \ldots m) \in S_m$. For a closed loop in $\Gamma(x)$, the line segments that define $\pi(x)$ and the ticks $\rightangle$ that define $e$ or $\xi$ must together close into a cycle in the product permutation $e \circ \pi(x)$ (respectively $\xi \circ \pi(x)$). We highlight this fact for future use:
\begin{equation}
\textrm{loops of $\Gamma(x)$} \quad \leftrightarrow \quad \textrm{cycles of $\pi(x)$ or $\xi \circ \pi(x)$}
\end{equation}
In particular, the number of loops in $\Gamma(x)$ is read off by the cycle-counting function $l$, that is $l(\pi(x))$ or $l(\xi \circ \pi(x))$. 

We close this paragraph with a theorem taken from \cite{stanley}, which will be useful in our proof. If a permutation $\pi \in S_m$ is noncrossing then:
\begin{equation}
l(\pi) + l(\xi \circ \pi) = m +1
\label{cyclecountsum}
\end{equation}
Therefore, $m+1$ is the total number of loops in $\Gamma(x)$ and $\Gamma(x')$ that cross the polar circle, where $x$ and $x'$ are related by flipping the $S_{A_1 A_2 \ldots A_m}$ bit. 

\medskip
{\bf Representing $\Gamma(x)$ and $\pi(x)$ on the plane}~~It is helpful to depict the loops of $\Gamma(x)$ on the plane. To do so, find a noncontractible cycle $c \subset \mathbb{RP}^2$, which intersects precisely one loop of $\Gamma(x)$ exactly once. (Exactly one intersection is necessary and guaranteed because precisely one loop of $\Gamma(x)$ is noncontractible.) Any additional intersections between $c$ and $\Gamma(x)$ are removable by homotopy; we exploit this freedom to find a cycle $c$ with the desired property. Now cut $c$ open and glue in a disk $D$; this effectively replaces a cross-cap with a disk and produces a sphere. We would like to visualize the loops of $\Gamma(x)$ on that sphere. The loops that do not intersect $c$ in $\mathbb{RP}^2$ (that is, all except one) are unaffected by the cutting and gluing we performed, so they can be drawn on the sphere directly. For the loop that does intersect $c$, its unique point of intersection lifts to two points on the sphere because after glueing the disk the loop must enter and exit $D$. For that loop, too, its lift to the sphere is topologically unambiguous. 

We now project the sphere stereographically onto the plane, such that the North/South pole---the boundary of the tiling in Figure~\ref{fig:rp2proof}---gets mapped to infinity. Several examples are shown in Figure~\ref{fig:rp2cases}. The polar circle, which is really a $2m$-gon, is intersected by $l(\pi)$ or $l(\xi \circ \pi)$ loops, as the case may be. Wherever a loop crosses the polar circle, it always crosses back at the nearest possible site; this is the fact that $\Gamma(x)$ only contains ticks $\rightangle$ on the boundary diagonal. One other feature of the stereographic projection is the disk $D$, which we glued onto the cut cycle $c$. As explained in the previous paragraph, the one noncontractible loop is distinguished from the other loops in $\Gamma(x)$ by the fact that its lift to the sphere intersects $D$. 

In summary, we represent the loops of $\Gamma(x)$ as a diagram on a plane. The diagram shows intersections of loops with the polar circle and indicates which loop is noncontractible. With the exception of the disk $D$, all other connected regions in the diagram faithfully represent the connected components, into which $\Gamma(x)$ divides $\mathbb{RP}^2$. Figure~\ref{fig:rp2cases} shows examples.

Consider the exterior of the polar circle in our diagram. That region comes from the `polar zone' in $\mathbb{RP}^2$, that is the boundary diagonal in Figure~\ref{fig:rp2proof}. In particular, it contains no vertices; the last line of vertices in Figure~\ref{fig:rp2proof} live on the polar circle itself. Because the polar zone contains no vertices of its own, it only affects the map $f$ by setting the connectivity between regions. It does so in a very regular way. 

To facilitate a definition, let us temporarily treat the polar circle on equal footing with the loops of $\Gamma(x)$. That is, we temporarily use both the loops of $\Gamma(x)$ and the polar circle to divide the plane---and, by extension, the projective plane---into connected components. Doing so produces a finer division of the [projective] plane than does $\Gamma(x)$ alone. In this finer division, the interior regions which are incident to the polar circle can be sorted into two classes. Because this classification mirrors the labeling of polar circle edges as odd and even, we likewise refer to the polar circle-reaching regions as odd and even.  

We now return to analyzing how $\Gamma(x)$ and $\Gamma(x')$ divide the [projective] plane into connected components. That is, from here on the polar circle is no longer understood to divide regions. 
Observe that with each value of the $S_{A_1 A_2 \ldots A_m}$ bit in $x$, either the odd or the even polar circle-reaching regions are connected through the polar zone. Conversely, the other class of circle-reaching regions are all cut off from each other by the little pieces of $\Gamma(x)$, which penetrate past the polar circle. (They are the ticks $\rightangle$ on the boundary diagonal in Figures~\ref{fig:rp2proof} and \ref{fig:rp2example}.) When the $S_{A_1 A_2 \ldots A_m}$ bit flips, the two classes exchange roles. That is, if $\Gamma(x)$ leaves the odd regions connected but splits up the even regions then $\Gamma(x')$ joins up the even regions but splits up the odd ones. 

Finally, we remark that the two classes of regions are in one-to-one correspondence with the cycles of $\pi$ and of $\xi \circ \pi$. Therefore, by equation~(\ref{cyclecountsum}), we always have a total of $m+1$ polar circle-reaching regions to play with. Either $l(\pi)$ of them are connected and $m+1-l(\pi)$ are all disjoint or the other way around.

\subsubsection{Proof that $f$ is a contraction}
Based on lemma~(\ref{rp2simplification}), we consider flipping one bit, which is associated with a single term on the LHS side of (\ref{rp2asymm2}). We treat flips on interior squares (all with coefficient 1) and the flip on $S_{A_1 A_2 \ldots A_m}$ (with coefficient $m-1$) separately. Let us start with the latter case, for which we just developed an extensive machinery. 

\begin{figure}[t!]
    \centering
    $\begin{array}{lr}
    \includegraphics[width=0.47\linewidth]{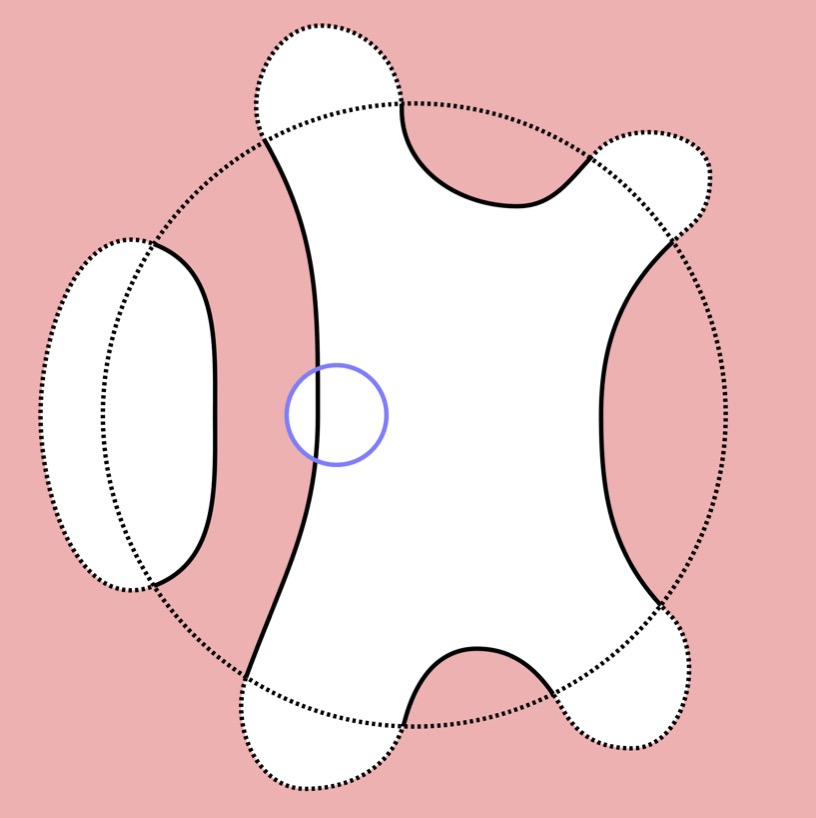} &
    \includegraphics[width=0.47\linewidth]{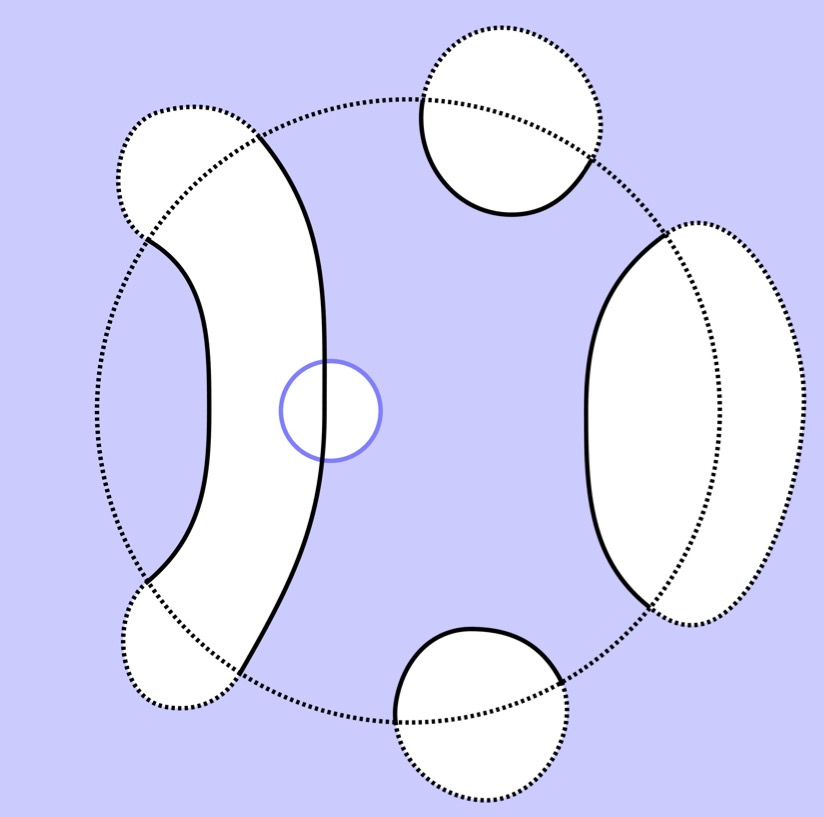} \\
    \includegraphics[width=0.47\linewidth]{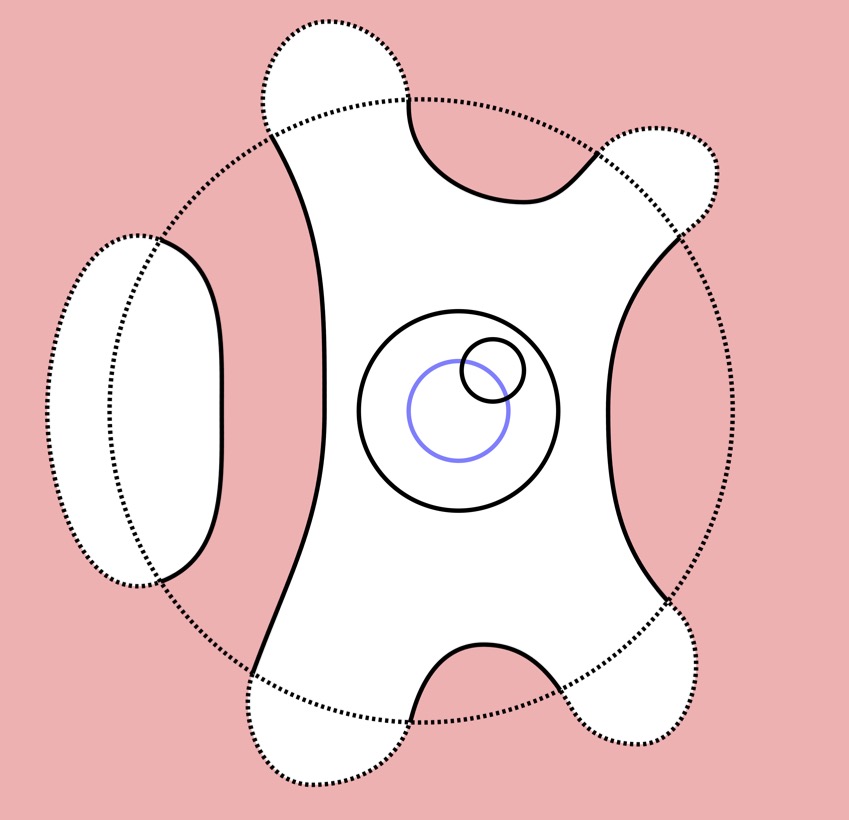} &
    \includegraphics[width=0.47\linewidth]{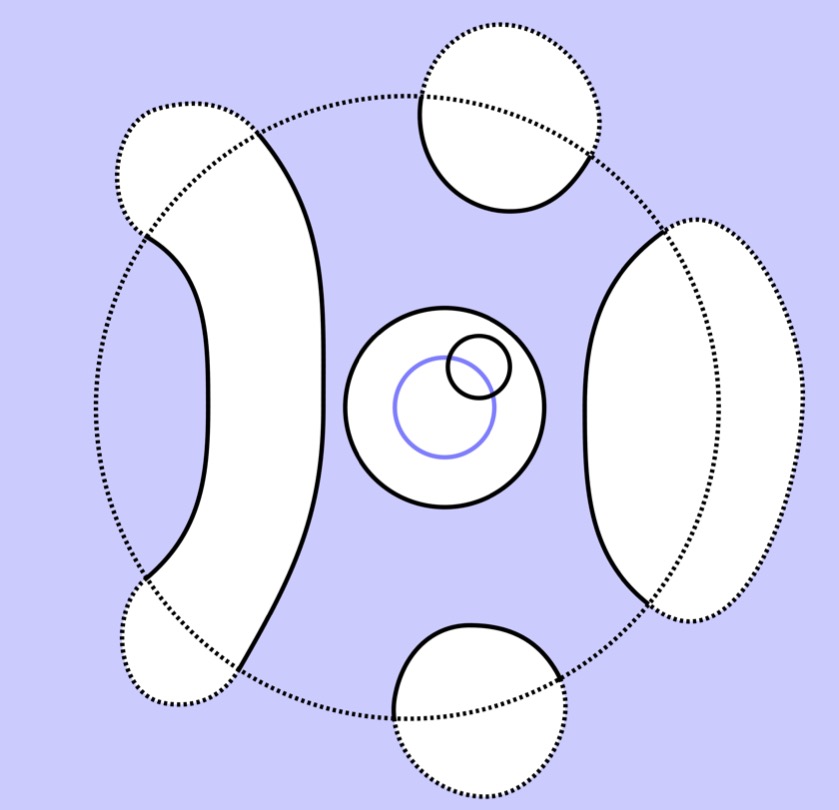} \\
    \includegraphics[width=0.47\linewidth]{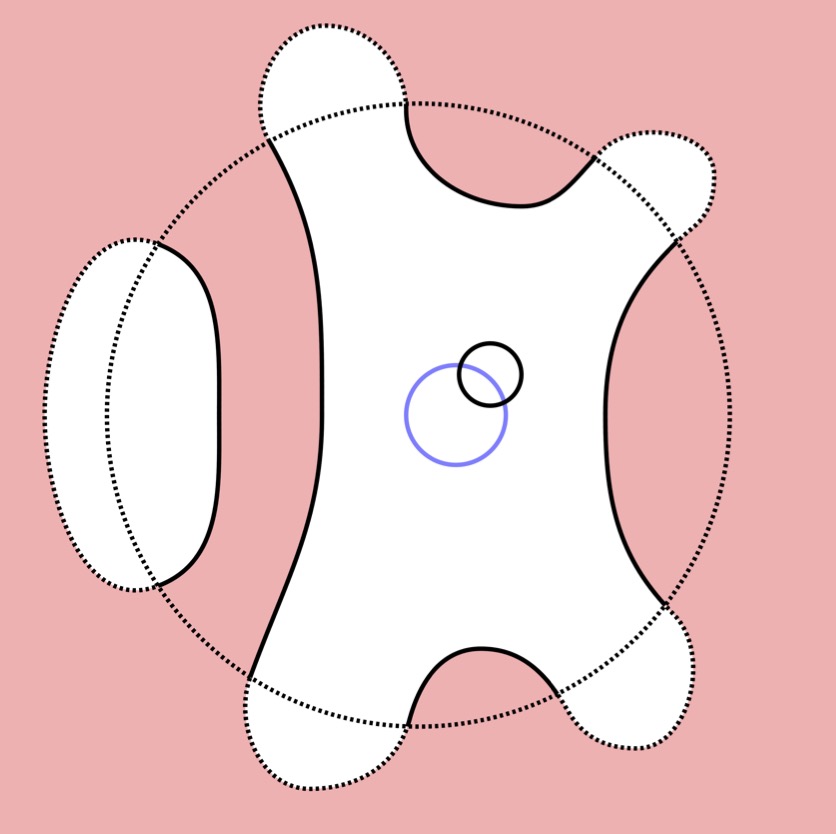} &
    \includegraphics[width=0.47\linewidth]{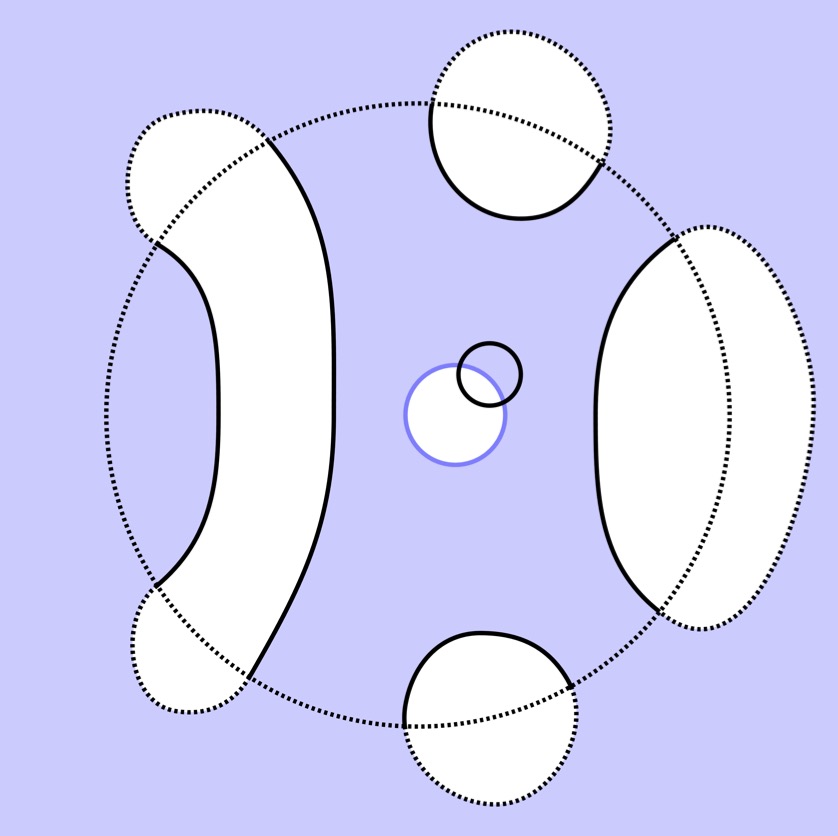}
    \end{array}$
    \caption{The loops of $\Gamma(x)$ on $\mathbb{RP}^2$, mapped diffeomorphically onto (plane $\setminus$ disk). The excised disk is inside the blue circle, which represents a cross-cap. The big circle is the polar circle; its exterior is the polar zone (boundary diagonal), which contains no vertices. There are three core scenarios (i-iii), depending on whether the cross-cap is intersected by a polar circle-reaching loop and, if not, whether some loop divides the cross-cap from the polar circle. In each scenario, we color the region that is connected via the polar zone, before and after flipping the $S_{A_1 A_2 \ldots A_m}$ bit in $x$.}
    \label{fig:rp2cases}
\end{figure}

\medskip
{\bf Flipping the $S_{A_1 A_2 \ldots A_m}$ bit is contracting}
We must show that flipping the $S_{A_1 A_2 \ldots A_m}$ bit alters at most $m-1$ entries in $f(x)$.  There are three cases to consider: (i) when the noncontractible loop in $\Gamma(x)$ crosses the polar circle and (ii-iii) when it does not. The latter cases are distinguished by whether there does (ii) or does not (iii) exist another contractible loop in $\Gamma(x)$, which separates the polar circle from the cross-cap. The three cases are shown in Figure~\ref{fig:rp2cases}. As explicit examples, Figure~\ref{fig:rp2bc} exemplifies case~(i) while Figure~\ref{fig:rp2example} exemplifies case~(iii).

In each case, we have $l(\pi)$ regions merging into one and one region splitting into $l(\xi \circ \pi)$ disjoint regions. To facilitate the discussion, we represent this operation in the notation:
\begin{align}
V_1, V_2,\, \ldots, V_{l(\pi)} & \,\, \to \,\, V = \cup_{j=1}^{l(\pi)}\, V_j \\
\cup_{j=1}^{l(\xi \circ \pi)}\, W_j = W & \,\, \to \,\, W_1, W_2,\, \ldots, W_{l(\xi \circ \pi)} \nonumber
\end{align}
The uppercase symbols represent collections of vertices $v_i$. Map $f(x)$ assigns 1 to the vertex $v_i$ with the smallest index $i$ in each collection, assuming that collection comes from a region that does not wrap a noncontractible cycle in $\mathbb{RP}^2$. We remind the reader that the vertex with the smallest index is called `top priority.' All other vertices are assigned 0.

A handy thing to observe is that the top priority vertex in a union of regions must also be top priority in one of the constituent regions. Therefore, when we adjoin $l(\pi)-1$ unwrapped regions $V_i$ to $V_1$, precisely $l(\pi)-1$ terms in $f(x)$ flip. To be explicit, we spell out two scenarios:
\begin{itemize}
\item Scenario 1: $V_1$ is wrapped (contains a noncontractible cycle in $\mathbb{RP}^2$). Initially, we have $l(\pi)-1$ unwrapped regions and so $f(x)$ contains $l(\pi)-1$ ones on $V = \cup_i V_i$. After the flip, $V$ is wrapped and all vertices in it are assigned zero.
\item Scenario 2: $V_1$ is unwrapped (does not contain a noncontractible cycle in $\mathbb{RP}^2$). Initially, we have $l(\pi)$ unwrapped regions and so $f(x)$ contains $l(\pi)$ ones on $V = \cup_i V_i$. After the flip, $V$ is unwrapped and only the top priority vertex in $V$ gets assigned a one. That vertex is top priority in one of the constituent regions being merged, so it was also assigned one before the flip.
\end{itemize}

We now discuss the three cases in turn; viz. Figure~\ref{fig:rp2cases}:
\begin{itemize}
\item Case~(i): One region $V_j$ is wrapped and one region $W_j$ is wrapped. Thus, the flip runs Scenario~1 on the $V_j$'s and runs Scenario~1 backwards on the $W_j$'s.
\item Case~(ii): All regions $V_j$ and $W_j$ are unwrapped. The flip runs Scenario~2 on the $V_j$'s and runs Scenario~2 backwards on the $W_j$'s.
\item Case~(iii): One region $V_j$ is wrapped and all regions $W_j$ are unwrapped. The flip runs Scenario~1 on the $V_j$s and runs Scneario~2 backwards on the $W_j$'s.
\end{itemize}
In all three cases, the number of altered bits in $f(x)$ is
\begin{equation}
\big(l(\pi) - 1\big) + \big(l(\xi \circ \pi) - 1\big) = m-1,
\end{equation}
where we use equation~(\ref{cyclecountsum}). 

\medskip
{\bf Flipping bits on interior squares} Finally, we must also show that a flip on an interior square alters at most one entry in $f(x)$. 

By the corollary highlighted in one of the preceding paragraphs, every bit flip on an interior square takes two loops into one (or vice versa). Therefore, all flips on interior squares are captured by list~(\ref{3effects}). Moreover, the option $1 + 1 \leftrightarrow 0$ cannot occur because two noncontractible loops cannot coexist in $\Gamma(x)$. In the end, we only need to consider bit flips, which rearrange the loops of $\Gamma(x)$ as $0 + 0 \leftrightarrow 0$ and $0 +1 \leftrightarrow 1$. There the argument is identical to Figure~\ref{fig:joiningloops}. The only real modification is to place appropriate arrows on the sides of the diamond so as to make a projective plane instead of the torus. In comparison with the toric argument, we also replace the designation `bottommost and rightmost vertex' with `top priority vertex,' but this change is immaterial for the argument. 

This concludes the proof of inequalities~(\ref{app:rp2ineqs}).
\end{document}